\documentclass[a4paper,11pt]{article}
\usepackage{amssymb}
\usepackage{amsfonts}
\usepackage{latexsym}

\def\beq{\begin{eqnarray}}
\def\eeq{\end{eqnarray}}
\def\ln{\,\mbox{ln}\,}
\def\Det{\,\mbox{Det}\,}

\def\tr{\,\mbox{tr}\,}
\def\Tr{\,\mbox{Tr}\,}

\def\al{\alpha}
\def\be{\beta}
\def\ch{\chi}
\def\ga{\gamma}\def\de{\delta}

\def\ep{\epsilon}
\def\ze{\zeta}

\def\ka{\kappa}
\def\la{\lambda}
\def\na{\nabla}

\def\rh{\rho}
\def\si{\sigma}
\def\om{\omega}

\def\ta{\tau}
\def\th{\theta}

\def\Ga{\Gamma}
\def\De{\Delta}
\def\La{\Lambda}
\def\Si{\Sigma}
\def\Om{\Omega}

\input{psfig}

\begin{document}

%
\begin{center}
{\large\sc Higher Derivative Quantum Gravity
with Gauss-Bonnet Term
}
\vskip 6mm
{\small \bf Guilherme de Berredo-Peixoto}
 \footnote{E-mail address: guilherme@fisica.ufjf.br}
\quad and \quad {\small \bf Ilya L. Shapiro} \footnote{On leave
from Tomsk State Pedagogical University, Russia. E-mail address:
shapiro@fisica.ufjf.br}
\vskip 6mm
{\small\sl Departamento de F\'{\i}sica -- ICE,
Universidade Federal de Juiz de Fora,} \\
{\small\sl Juiz de Fora, CEP: 36036-330, MG,  Brazil}
\end{center}
\vskip 6mm
\centerline{\uppercase{abstract}}
\vskip 2mm

\begin{quotation}
\noindent Higher derivative theory is one of the important models
of quantum gravity, renormalizable and asymptotically free within
the standard perturbative approach. We consider the $4-\epsilon$
renormalization group for this theory, an approach which proved
fruitful in $2-\epsilon$ models. A consistent formulation in
dimension $n=4-\epsilon$ requires taking quantum effects of the
topological term into account, hence we perform calculation which
is more general than the ones done before. In the special $n=4$
case we confirm a known result by Fradkin-Tseytlin and
Avramidi-Barvinsky, while contributions from topological
term do cancel. In the more general case of $4-\epsilon$
renormalization group equations there is an extensive ambiguity
related to gauge-fixing dependence. As a result, physical
interpretation of these equations is not universal unlike we
treat $\epsilon$ as a small parameter. In the sector of
essential couplings one can find a number of new fixed points,
some of them have no analogs in the $n=4$ case.
\vskip 6mm

\end{quotation}

\vskip 6mm

\section{\large\bf  Introduction}

Quantizing gravitational field is an important aspect of
quantum description of nature. Besides main efforts in the
last decades were concentrated on the study of (super)string
theories, recently there was an extensive development of a proper
quantum gravity: both perturbative \cite{hove,dene,stelle,frts82}
and non-perturbative \cite{Weinberg79,ambjorn,smolin,reuter}. It
is well known that the perturbative approach meets serious
difficulty which can be summarized as a conflict between
renormalizability and unitarity \cite{Weinberg79}.
Quantum gravity based on General Relativity is non-renormalizable,
in particular higher derivative divergences emerge starting
already from the one-loop level \cite{hove,dene}. In the presence of
matter fields or at second loop \cite{GorSag} the divergences
persist on-shell, indicating that no ``magic'' cancellations
occur. The situation in supergravity extensions of General
Relativity is somehow better, because higher derivative
divergences emerge only at higher loops \cite{west}. However, from
the general perspective there is no principal difference, for the
corresponding version of supergravity extensions of General
relativity are also non-renormalizable by power counting.

An alternative way of quantizing gravity is to introduce some
higher derivative terms into a classical action, treating them
at the same footing as the lower-derivative (Einstein-Hilbert and
cosmological) terms. For example, adding generic fourth order
derivative terms, one modifies propagator and vertices in such
a way that
the new quantum theory is renormalizable \cite{stelle} (see also
\cite{votyu84} for a more general formal proof of
renormalizability). This nice property enables one to establish
the asymptotic freedom in the UV limit
\cite{julton,salstr,frts82,avrbar} and explore the possible role
of quantum gravity in the asymptotic behavior for GUT-like
models \cite{bush}. Introducing new terms with derivatives higher
than four leads to super-renormalizable theories of quantum
gravity \cite{highderi}.

Unfortunately, the price for renormalizability is very high.
The propagator of renormalizable quantum gravity contains
unphysical poles corresponding to the states with negative kinetic
energy or with a negative norm in the state space. These
unphysical states are conventionally called massive ghosts
\cite{stelle} \footnote{General introduction to the fourth order
derivative quantum gravity can be found in \cite{book}. A detailed
investigation of propagator in higher derivative gravity models
has been recently performed in \cite{accioly} (see also further
references therein).}. Despite the ghost masses have magnitude
of the Planck order, they spoil unitarity of the S-matrix exactly
in the range of energies where quantum gravity is supposed to
apply\footnote{There are interesting examples of higher derivative
gravity theory with torsion which are free of ghosts
\cite{nevill,sezgin}, but renormalizability can not be
achieved in these models \cite{tyutin,torsi}.}. Indeed the
presence of massive unphysical ghosts in the tree-level propagator
does not necessary imply that S-matrix is not unitary if one
takes quantum corrections into account. It has been readily
noticed \cite{salstr,tom} that the quantum corrections to
graviton propagator may, in principle, transform the real
unphysical pole into a pair of complex poles. In this case the
massive states should disappear in the asymptotic states and hence
unitarity of the S-matrix gets restored. Unfortunately, until
now there is no safe method to check whether this is really the
effect of a resummation in the perturbative expansion. For
example, despite the leading order of $\,1/N\,$ expansion
provides corrections in appropriate form \cite{tom,anto}, this
approximation is not consistent in the case of quantum gravity
\cite{dgo}. Qualitatively the same is also true for other attempts
in this direction and the final answer concerning the presence of
ghosts in a full quantum theory remains unclear. In generic
higher derivative gravitational theory (polynomial in derivatives)
we meet complicated structure of the propagator, but with
inevitable appearance of unphysical massive ghosts
 \cite{highderi}.

In order to clarify the situation with massive ghosts one needs to
perform a non-perturbative treatment of the theory. In quantum
theory, one of the known non-perturbative methods is the
consideration of theory in a dimension $\,4-\ep\,$, where
$\,\ep\,$ is a small parameter. This method has
been developed by Wilson and Fisher for investigating critical
phenomena \cite{Wilson71} and later on applied to quantum
field theory problems \cite{Wilson73}. In quantum gravity the
same idea has been consistently applied only in two-dimensional
models \cite{GKT,ChrDuff,Weinberg79,kawai,sakai}, where it proved
very fruitful. The main lesson of the two-dimensional case is
the correspondence between $\,2+\ep\,$ approach and numerical
non-perturbative methods, such as dynamical triangulations
\cite{ambjorn}. One can accept this is a hint indicating the
potential importance of the $\,4-\ep\,$ consideration. In the
present paper we shall consider the $\,4-\ep\,$ approach for
higher derivative quantum gravity. In particular, it looks
interesting to explore the possibility of new fixed points in
$\,4-\ep\,$ theory. The important aspect is that the Gauss-Bonnet
term, which is a part of higher derivative gravity action, is
quite relevant for this purpose.
The reason is that, for the theory
in $\,n = 4-\ep\,$ dimensions, the Gauss-Bonnet term is not
a topological term, in particular it contributes to
many vertices and may be, in principle, affecting the
$\be$-functions in the $\,4-\ep\,$ case.

The next question is whether the effect of the Gauss-Bonnet term
has to vanish in $\,n\to 4\,$ limit. In fact, the opposite
situation can not be ruled out beforehand \cite{capkim}. The point
is that the topological nature of this term is closely related to
its diffeomorphism invariance. However, when theory is
quantized (e.g., through the Faddeev-Popov procedure), the
covariance is broken such that the vector space extends beyond
physical degrees of freedom. After quantization, not only spin-2,
but also spin-1 and spin-0 components of quantum metric become
relevant, and the topological term creates new vertices of
interaction between these components \cite{capkim}. The
quantum-gravitational loops may be, in principle, affected by the
presence of topological term. Of course, this output does not
look probable because, after including the topological term, the
gauge-fixing condition should modify and eventually compensate the
new vertices. But this is a sort of believes which are always good
to verify and we will perform such verification in what follows.
In a recent paper \cite{Weyl} we performed similar calculation
for the particular case of Weyl quantum gravity and found that, in
this case, the effect of Gauss-Bonnet term is relevant for
\ $\,4-\ep\,$ renormalization group but vanish when $\,n\to 4$. An
extra benefit of the calculational scheme taking the
Gauss-Bonnet term into account is the highest degree of safety for
correctness of calculations. In case of Weyl theory we
have confirmed the previous results \cite{frts82,antmot} for the
$\,n = 4\,$ renormalization group and also established
conformal invariance of one-loop counterterms. Here we shall
generalize this calculation for generic non-conformal theory
and present a more complete investigation of \ $\,4-\ep\,$
renormalization group, taking into account a gauge-fixing
dependence and the corresponding ambiguity.

In order to complete the picture and emphasize even more
the relevance of the problem, let us remember that the
Gauss-Bonnet-like term is an important ingredient of an
effective low-energy action of the (super)string theory.
The string effective action is composed by the Einstein-Hilbert
term and an
infinite set of higher derivative terms which emerge in the
next order in $\,\al^\prime$ (see, e.g., \cite{GSW}).
Therefore, models like effective low-energy quantum gravity
\cite{don,tsamis} which are supposed to be a kind of some
universal low energy limit of string theories, may be sensitive
to the presence of Gauss-Bonnet term. The uniqueness of the
output in effective approaches
depends on whether the $\,n\to 4\,$ limit is universal
or not. Hence in case we find vanishing contribution from
the Gauss-Bonnet term in
$\,n\to 4\,$ limit, this will be a certain confirmation of
universality of quantum corrections.

Thus, the main purpose of our work is to derive the one-loop
corrections for higher-derivative quantum gravity with the
Gauss-Bonnet term in $\,n\neq 4$ and analyze the consequent
renormalization group. The paper is organized as follows. In the
next section we introduce notations and briefly discuss
the theory. Section 3 contains brief summary of a Lagrangian
quantization of the theory and in particular the discussion of
gauge-fixing conditions and related arbitrariness in
counterterms. In section 4 we derive the trace of the coincidence
limit of second Schwinger-DeWitt coefficient for an arbitrary
dimension $n$. Furthermore, we consider the $\,n\to 4\,$ limit and
obtain the expression for the divergences. The cancellation of
contributions from topological term and perfect fitting with
the previous calculation of divergences in the higher derivative
quantum gravity \cite{avrbar} confirm the correctness of our
calculation. In section 5 we perform analytical and numerical
investigation of $\,4-\ep\,$  renormalization group equations
for the higher derivative couplings and in section 6 consider
the renormalization group equations for the Newton and
cosmological constants.
Finally, in section 7 we draw our conclusions. Many technical
details and bulky formulas are collected in Appendices.
Throughout the paper we use Euclidean signature in order to be
consistent with the previous publications \cite{frts82,avrbar}.

\section{\large\bf  General framework and notations}

The classical action of the theory has the form
\beq
S\,=\,-\,\mu^{n-4}\, \int d^nx\sqrt{g}\,
\Big\{\,\,\frac{1}{2\la}\,C^2 \,-\,\frac{1}{\rho}\,E
\,+\,\frac{1}{\xi}\, R^2 \,+\,\tau\,\square R -
\frac{1}{\kappa^2}\, (R - 2 \La )\,\Big\}\,, \label{action}
\eeq
where \ $\mu$ \ is some dimensional parameter,
\beq
C^2\,=\,R_{\mu\nu\al\be}R^{\mu\nu\al\be}
\,-\,\frac{4}{n-2}\,R_{\al\be}R^{\al\be}
\,+\,\frac{2}{(n-1)(n-2)}\,R^2
\nonumber
\eeq
is the square of Weyl tensor,
\beq
E\,=\,R_{\mu\nu\al\be}R^{\mu\nu\al\be}
\,-\,4\,R_{\al\be}R^{\al\be}\,+\,R^2
\nonumber
\eeq
is the integrand of the Gauss-Bonnet term, which is
topological in $n=4$ and $\,\la,\,\rho,\,\xi,\,\tau\,$ \ are
independent parameters in the higher derivative sector of the
action. Furthermore, $\Box = g^{\mu\nu}\na_\mu\na_\nu$ and
$\,\ka^2= 16\pi G$, where $\,G\,$ is the
Newton constant and \ $\La$ \ is the cosmological constant.

An alternative form for the action (\ref{action}) is
\beq
S =
-\mu^{(n-4)}\int d^n x\sqrt{g}\, \left\{\, x\, R_{\mu\nu\al\be}^2
\,+\,y\, R_{\mu\nu}^2 \, + z\,R^2\,+\,\tau\square R\,-\,
\frac{1}{\kappa ^2}\, (R - 2 \La )\, \right\}\,, \label{higher}
\eeq
where we denote \ $R_{\mu\nu\al\be}^2 = R_{\mu\nu\al\be}
R^{\mu\nu\al\be}$ \ and \ $R_{\al\be}^2 = R_{\al\be}R^{\al\be}$.
The parameters \ $x$, $y$, $z$ \ are related to
$\,\rho$, $\,\la$, $\,\xi\,$ as follows:
\beq
x =
\frac{1}{2\,\lambda}\,-\,\frac{1}{\rho}
\,,\qquad
y \,=\,-\,{\frac{2}{\left (n-2\right )\lambda}} +\frac{4}{\rho}
\,,\qquad
z \,=\, -\frac{1}{\rho}
+ \frac{1}{\xi} + {\frac{1}{\la\,(n-1)(n-2)}} \,.
\label{x y z}
\eeq
The two forms of the action (\ref{action}) and (\ref{higher})
are completely equivalent and we shall use one or another
depending on convenience.

At the classical level in the four dimensional space the role
of the Gauss-Bonnet $\,\int E\,$-term is negligible. However,
situation changes when theory is considered in $\,n\neq 4$,
where the
$\,\int E\,$-term becomes dynamical and hence must be taken into
account. Using the language of Feynman diagrams, one can say that
despite this term does not contribute to graviton propagator
on flat background for any \ $n$, it does contribute to vertices
of interaction between different modes of metric (spin-2, spin-1
and spin-0). Hence the trivial nature of this term must be taken
with proper caution, especially in $\,n\neq 4$. Let us mention
that, for $\,n = 4-\ep$, there is a natural limit $\,\ep=1\,$
imposed by the following circumstance. For $n=3$ the expressions
for $E$ and $C^2$ coincide, and since the Weyl tensor vanish at
$n=3$, the Gauss-Bonnet term also vanish.

At quantum level, all terms in the action (\ref{action}) are
necessary for renormalizability. At the same time, the term
\ $C^2$ has a key role. If we do not include this term, the
quantum theory will have, in the graviton (traceless and
transverse) sector of quantum metric the situation when the
propagator has only second derivatives while there are four
derivative vertices. Then the standard evaluation of the
superficial degree of divergences for the diagrams of the theory
(see, e.g. \cite{book}) shows very bad renormalizability
properties which are even worse than the ones for the quantum
gravity based on General Relativity. On the other hand, since
the term \ $C^2$ \ is included, the propagator has,
in the momentum representation, the form
\beq
G\,\,\sim\,\,\frac{1}{k^2\,(k^2+m^2)}
\,\,=\,\, \frac{1}{m_2^2}
\,\Big(\,\frac{1}{k^2}-\frac{1}{k^2+m_2^2}\,\Big)\,,
\label{ghost}
\eeq
where the negative sign of the second term indicates the ghost
nature of the propagating spin-2 state with mass $\,m_2$.
Other unphysical mode is possible in the spin-0 sector of
metric, depending on the coefficient of the \ $\int R^2$-term
\cite{stelle,book,accioly}.

Without \ $R^2$-term, the higher derivative sector of
theory possesses (in \ $n=4$ \ case) local conformal
invariance and its renormalization has some special features. The
main point is that conformal symmetry is violated by anomaly,
therefore this symmetry is not supposed to hold in
renormalization beyond one-loop level. We have recently
discussed renormalization in conformal quantum gravity in
\cite{Weyl} and hence will not consider it here.

The relevance of the Gauss-Bonnet term at quantum level
is also significant. Long ago has been noticed
that the topological nature of this term is related
to Bianchi identity \cite{capkim}. As we already
mentioned in the Introduction, it is in principle
possible that the loop contributions depend on the
presence of topological term. In particular, there
is no reason to expect that the
effect of Gauss-Bonnet term is negligible for
$n=4-\ep$-dimension renormalization group, for this term
is not topological at $n\neq 4$ even at classical
level. In previous paper \cite{Weyl} we have
confirmed these considerations for the special case
of Weyl quantum gravity and will now generalize
these results for general higher derivative quantum
gravity (\ref{action}).

\section{\large\bf  Quantization and gauge fixing}

The general scheme of Lagrangian quantization for theory
(\ref{action}) has been formulated in \cite{stelle} (see also
\cite{book} for more detailed exposition). The most useful method
of practical calculations is through the background fields method
and Schwinger-DeWitt technique (see
\cite{bavi85} for an extensive introduction including the
higher derivative case). The application of these methods to the
case of higher derivative gravity has some peculiarities which
have been revealed in \cite{frts82} (see also \cite{Christen,book}
for more detailed pedagogical consideration).

The background field method implies special parametrization
of the metric
\beq
g_{\mu\nu}\,\,\longrightarrow\,\,g^\prime_{\mu\nu}
\,\,=\,\,g_{\mu\nu}+h_{\mu\nu}\,.
\label{background}
\eeq
In the {\it r.h.s.} of the last formula,
$\,g_{\mu\nu}\,$ is a background metric and $\,h_{\mu\nu}\,$ is
a quantum field (integration variable in the path integral). The
1-loop effective action  $\,\bar{\Ga}^{(1)}\,$ can be written as
\cite{frts82}
\beq
\bar{\Ga}^{(1)}[g_{\mu\nu}] \,=\,
\frac{i}{2}\,\ln\,\Det \hat{\cal{H}}\,\,-\,\,
\frac{i}{2}\,\ln\,\Det \,Y^{\al\be} \,\,-\,\, i\,\ln\,\Det
\hat{\cal{H}}_{gh}\,, \label{e}
\eeq
where $\,\hat{\cal{H}}\,$ is
bilinear (in the quantum fields)  form of the action
(\ref{action}), taken together with the gauge-fixing term
\beq
S_{GF}\,=\,\mu^{n-4\,}\int
d^nx\sqrt{g}\,\,\chi_\al\,Y^{\al\be}\,\chi_\be\,.
\label{gauge}
\eeq
$\hat{\cal{H}}_{gh}\,$ is bilinear form of the action of
the Faddeev-Popov ghosts, $\,\mu\,$ is the renormalization
parameter in dimensional regularization. The expression (\ref{e})
includes also the contribution of the weight function
$\,Y^{\al\be}$. In the case of higher derivative quantum
gravity theory this term gives relevant contribution to the
effective action, because $\,Y^{\al\be}\,$ is a second order
differential operator \cite{frts82}.

The general form of the gauge fixing conditions (here we
restrict our attention to linear background gauges) has the
form
\beq
\ch^\mu & = &
\na_{\la}h^{\la\mu}+\beta\,\na^\mu h\ \nonumber
\\
Y_{\mu\nu} & = & \frac{1}{\al}\,\Big( g_{\mu\nu}\Box +
\ga\na_\mu\na_\nu -\de\na_\nu\na_\mu + p_1R_{\mu\nu}
+p_2R\,g_{\mu\nu}\Big)\,,
\label{fixing}
\eeq
where
$\,\al_i\,=\,(\al,\,\be,\,\ga,\,\de,\,p_1,\,p_2)\,$ are arbitrary
gauge-fixing parameters. The action of the Faddeev-Popov ghosts
can be written as
\beq
S_{gh} \,=\, \mu^{n-4}\,\int d^n
x\sqrt{g}\,\,{\bar C}^\mu\, \left({\cal H}_{gh}\right)_\mu^\nu\,
C_\nu\,, \label{ghost action} \eeq where \beq \hat{\cal
H}_{gh}=\left({\cal H}_{gh}\right)_\mu^\nu = - \de_\mu^\nu\,\Box -
\na^\nu \na_\mu - 2\be \na_\mu \na^\nu\,. \label{ghost operator}
\eeq

Generally speaking, the counterterms may depend on the choice of
gauge-fixing parameters, but this dependence is constrained by
the on-shell gauge-fixing independence. Let us consider these
restrictions, generalizing similar discussion of \cite{shja} for
the general $\,n\,$ case.

We denote \ ${\Gamma}({\alpha}_{i})$ \ the effective action
corresponding to arbitrary values of gauge parameters \
${\alpha}_{i}$ \ and \ ${\Gamma}_m = {\Gamma}({\alpha}_{i}^{(0)})$
\ calculated for special values of these parameters, \
${\alpha}_{i}^{(0)}$. Our purpose is to evaluate the gauge
fixing dependence
\ ${\Gamma}({\alpha}_{i})$ \ or, equivalently, the expression for
the difference between the two effective actions
\ ${\Gamma}({\alpha}_{i})-{\Gamma}_m$. In general, this expression
may be rather complicated (see the details in \cite{shja}), but
the part of \ ${\Gamma}({\alpha}_{i})-{\Gamma}_m$ \
which is relevant for the renormalization group
may be easily evaluated without special calculations. For this end
we remember that the gauge dependence of counterterms has to
disappear on the classical mass-shell (see, e.g., \cite{VLT} for a
general proof for gauge theories). Hence we can write \beq
{\Gamma}({\alpha}_{i}) = {\Gamma}_m + \int {d^n}x \sqrt{g}\,\,
f_{\mu\nu} ({\alpha}_{i})\,\,\frac{\delta S}{\de g_{\mu\nu}}\,,
\label{a1} \eeq where $\,f_{\mu\nu} ({\alpha}_{i})\,$ is some
unknown function. The integration is taken over \ $n$-dimensional
space, because our target is the renormalization group in
$n=4-\ep$ dimensions.

Since the object of interested is the divergent part of the
effective
action \ ${\Gamma}({\alpha}_{i})$, one can assume it is a local
expression. Furthermore, both \ ${\Gamma}_m$ and
${\Gamma}({\alpha}_{i})$ have the same dimension as the classical
equations of motion \ ${\delta S/{\delta g_{\mu\nu}}}$. Therefore
\ $f_{\mu\nu} ({\alpha}_{i})$ \ is a symmetric dimensionless
tensor and the unique choice for it is
\beq
f_{\mu\nu}
({\alpha}_{i}) = g_{\mu\nu}\,\times\, f({\alpha}_{i})\,,
\label{a2}
\eeq
where $f({\alpha}_{i})$ is a numerical quantity.
Thus we arrive at the relation
\beq
{\Gamma}({\alpha}_{i}) &=&
{\Gamma}_m \,+\, f({\alpha}_{i})\,\times\, \int {d^n}x\, \sqrt
{g}\,\, g_{\mu\nu}\,\,{\delta S\over {\delta g_{\mu\nu}}}\,.
\label{a3}
\eeq
According to the last relation, the gauge-fixing
dependence of divergent part (or, more general, the local
part) of one-loop effective action is proportional to the
trace of classical equations of motion. The corresponding
traces for the relevant terms have the form
\beq
g_{\rho\si}{\delta \over {\delta g_{\rho\si}}} \,\int d^4x\,\sqrt
{g}\,R_{\rho\si\la\th}^2 &=& {\sqrt {g}}\,\left\{
\frac{n-4}{2}\,R_{\rho\si\la\th}^2 - 2\,{\Box}R \right\}\,,
\nonumber
\\
g_{\rho\si}{\delta \over {\delta g_{\rho\si}}} \,\int d^4x\,\sqrt
{g}\,R_{\rho\si}^2 &=& {\sqrt {g}}\,\left\{
\frac{n-4}{2}\,R_{\rho\si}^2 - \frac{n}{2}\,{\Box}R \right\}\,,
\nonumber
\\
g_{\rho\si}{\delta \over {\delta g_{\rho\si}}} \,\int d^4x\,\sqrt
{g}\,R^2 &=& {\sqrt {g}}\,\left\{ \frac{n-4}{2}\,R^2 -
2(n-1)\,{\Box}R \right\}\,, \nonumber
\\
g_{\rho\si}{\delta \over {\delta g_{\rho\si}}} \,\int d^4x\,\sqrt
{g}\,\big(\,R-2\La\,\big) &=& {\sqrt {g}}\,\left\{
\frac{n-2}{2}\,R - n\,\La \right\}\,. \label{shift 4}
\eeq

At this point it is better to separate the $n=4$ and $n=4-\ep$
cases. For $n=4$ we obtain the following formula for
the gauge-fixing dependence \cite{shja}: \beq
\Gamma(\alpha_{i})=\Gamma_m\,+\, f({\alpha}_{i})\,\times\, \int
d^4x\sqrt{g} \,\Big\{\,- \frac{1}{\kappa^2} \, (R-4\Lambda)
\,+\,\frac{6}{\xi}\,\Box R\,\Big\}\,. \label{total 4} \eeq

According to (\ref{total 4}), the coefficients of $E$,
$C^2$ and $R^2$ poles are gauge-fixing invariant. One of the
consequences is that the corresponding renormalization group
equations are universal, providing information about the UV limit
of the theory. At the same time, \ $(\Box R)$-type pole depends
on the choice of the gauge-fixing condition.
Taking different values of gauge-fixing
parameters $\,\al_i\,$ one can provide any desirable value of the
function $\,f(\al_i)\,$ and hence change the \ $(\Box R)$-type
counterterm. Thus, the corresponding parameter \ $\tau$ \ in
(\ref{action}) is not essential. The immediate conclusion is that
there is no much interest to calculate the renormalization of \
$\tau$, especially if the calculation is done for a particular
gauge-fixing. Another situation takes place for the
Einstein-Hilbert and cosmological terms. Renormalization of
each of them is gauge-fixing dependent, but one can easily check
that the dimensionless combination \ $\kappa^2\Lambda$ is an
essential coupling constant with the invariant renormalization
relation.

In the $n=4-\ep$ case the situation is more complicated. Simple
calculation using (\ref{shift 4}) yields the following
generalization of (\ref{total 4}):
\beq
{\Gamma}({\alpha}_{i}) &=&
{\Gamma}_m \,+\, f({\alpha}_{i})\,\times\,\mu^{n-4}\, \int {d^n}x
\sqrt{g}\, \Big\{\,\Big[2x+\frac{n y}{2}+2(n-1)z\Big]\,(\Box R)
\nonumber
\\
&+& \frac{n-4}{2}\,\big( x\,R_{\mu\nu\al\be}^2 \,+\,y\,
R_{\mu\nu}^2 \, + z\,R^2\big) \,-\, \frac{n-2}{2\,\ka^2}\,R \,+\,
\frac{n\,\Lambda}{\ka^2} \,\Big\}\,. \label{total n}
\eeq
An important consequence of the eq. (\ref{total n}) is that
neither one of
the parameters $\,x,y,z,\tau,\ka,\La$ is essential in the
$n=4-\ep$ case. However, gauge-fixing dependence is
concentrated in a single numerical function $f(\al_i)$ and
therefore we can easily extract combinations of the couplings
which are essential parameters. This result will be extensively
used in sections 5 and 6, where we consider renormalization
group equations for essential couplings.

\section{\large\bf
Bilinear expansion and derivation of divergences}

In order to apply the background field method, we need the
bilinear expansion in $\,h_{\mu\nu}\,$ for the classical
action (\ref{action}). It proves useful to work with the
equivalent form of the action (\ref{higher}), and therefore
we expand this action as follows:
\beq
S^{(2)} & = & -\mu^{(n-4)}\int d^n x\,\Big\{
      x\big[\sqrt{g}\,R^2_{\mu\nu\al\be}\big]^{(2)}
+ y\big[\sqrt{g}\,R^2_{\mu\nu}\big]^{(2)} \,+\,
z\big[\sqrt{g}\,R^2\big]^{(2)}
\nonumber
\\
& - & \frac{1}{\ka^2}\,
\left[\,\sqrt{g}\,R-2\sqrt{g}\,\La\,\right]^{(2)}\Big\}\,.
\label{expansion}
\eeq
The expressions
$\big[\sqrt{g}R^2_{\mu\nu\al\be}\big]^{(2)}$, \
$\big[\sqrt{g}R^2_{\mu\nu}\big]^{(2)}$, \
$\big[\sqrt{g}R^2\big]^{(2)}$, \ $\left[\sqrt{g}R\right]^{(2)}$,
$\left[\sqrt{g}\,\La\,\right]^{(2)}$ were derived in the previous
work \cite{Weyl} and we will not reproduce them here\footnote{All
but the first expansions can be also found in
\cite{julton,ChrDuff,avrbar,book}.}. Using these formulas and the
expression for the gauge-fixing term (\ref{gauge}), after
performing some cumbersome commutations \cite{Weyl} one can find
the bilinear form of the action
\beq \left[\,
S+S_{GF}\,\right]^{(2)} &=& h^{\mu\nu}\, {\cal
H}_{\mu\nu,\,\al\be}\, h^{\al\be}\,. \label{bilinearS}
\eeq
The operator $\hat{{\cal H}} = {\cal H}_{\mu\nu ,\,\al\be}$ depends
on gauge parameters, $\al$, $\be$, $\ga$, $\de$, $p_1$ and $p_2$.
For practical calculations these parameters have to be chosen in
such a way that ${\cal H}$ assumes the most simple (minimal) form
\beq
\hat{\cal{H}}\,=\,{\hat
K}\,\square^2+{\cal O}(\na^2)\,, \label{minimal high}
\eeq
where
$\,{\hat K}\,$ is a non-degenerate $c$-number operator. The
expressions for the gauge-fixing parameters providing cancellation
of the non-minimal four derivative terms
$\,g_{\mu\nu}\na_\al\Box\na_\be\,$,
$\,g_{\al\be}\na_\mu\Box\na_\nu\,$,
$\,\na_\mu\na_\nu\na_\al\na_\be\,$ and
$\,g_{\nu\be}\na_\mu\Box\na_\al\,$ are the following:
\beq
\beta =
\frac{y+4z}{4(x-z)}\,,\qquad \al = \frac{2}{y+4x}\,, \qquad \ga =
\frac{2x-2z}{y+4x}\,, \qquad \de=1\,,\qquad p_1=p_2=0\,.
\label{minimal}
\eeq
With this choice of the gauge parameters, the
operator $\hat{\cal H}$ takes the form (\ref{minimal
high})
\beq
\hat{\cal H} \,=\,\hat{K}\Box
^2+\hat{D}^{\rh\la}\na_\rh\na_\la + \hat{N}^\la\na_\la -(\na_\la
\hat{Z}^\la)+\hat{W}\,, \label{bilinear}
\eeq
where $\,\hat{K}$,
$\,\hat{D}^{\rh\la}$, $\,\hat{N}^\la$, $\,\hat{Z}^\la\,$ and
$\,\hat{W}\,$ are local matrix expressions in the
$\,h^{\mu\nu}$-space. The identity matrix in this space is defined
as a symmetric tensor
\beq
\hat{1}\,=\,\de_{\mu\nu ,\,\al\be}
\,=\, \frac12\,\left( g_{\mu\al}\,g_{\nu\be}\,+\,
g_{\mu\be}\,g_{\nu\al}\right)\,.
\eeq
Before writing down the
bulky formulas for the elements of (\ref{bilinear}), let us notice
that the expressions for $\,\hat{N}^\la\,$ and $\,\hat{Z}^\la\,$
are in fact irrelevant. The reason is that both of them are
proportional to the covariant derivatives of the curvature tensor.
Therefore, due to the locality of divergences, these terms may
contribute only to irrelevant gauge-fixing dependent
$\,\int\sqrt{g}\Box R$ counterterm. Since we are not
calculating this term here, for the sake of simplicity,  in what
follows we shall simply set both $\,\hat{N}^\la\,$ and
$\,\hat{Z}^\la\,$ to zero.

The matrix coefficient $\,\hat{K}$ of the fourth derivative term
has the following form:
\beq
(\hat{K})_{\mu\nu ,\al\be} & = &
\frac{y+4x}{4}\Big[\de_{\mu\nu
,\al\be}+\frac{y+4z}{4(x-z)}\,g_{\mu\nu}\,g_{\al\be}\Big]\,.
\label{K}
\eeq
The expressions for $\,\hat{D}^{\rh\la}$ and
$\,\hat{W}\,$ have the form
\beq
(\hat{D}^{\rh\la})_{\mu\nu
,\al\be} & = & - \,\,
2x\,g_{\nu\be}R_\al\mbox{}^{\rh\la}\mbox{}_\mu + 4x\,\de^\rh_\nu
R^\la\mbox{}_{\al\mu\be} + (3x+y)\,g^{\rh\la}R_{\mu\al\nu\be}
\nonumber
\\
& - &
(4x+2y)\,\de_\al^\rh g_{\nu\be}\,R^\la\mbox{}_\mu
- 2x\,g_{\nu\be}R_{\mu\al}g^{\rh\la}
+ y\,g_{\mu\nu}\de_\al^\rho R^\la\mbox{}_\be
\nonumber
\\
& - &
2z\,\de_{\al\be}\mbox{}^{\rh\la}R_{\mu\nu}
+ 2z\,g_{\mu\nu}g^{\rho\la}R_{\al\be}
+ 2x\,g_{\mu\nu}R_{\al}\mbox{}^{\rho\la}\mbox{}_{\be}
+ 2x \de_{\nu\be ,}\mbox{}^{\rh\la}R_{\mu\al}
\nonumber
\\
& + &
\big( \frac{z}{2}\,R\,-\,\frac{1}{4\ka^2}\Big)
\Big(
  \de_{\mu\nu ,\al\be}\,g^{\rh\la}
- g_{\mu\nu}\,g_{\al\be}\,g^{\rh\la}
- 2g_{\nu\be}\de_{\mu\al ,}\mbox{}^{\rh\la}
+ 2g_{\mu\nu}\de_{\al\be,}\mbox{}^{\rho\la}\big)
\nonumber
\\
& + &
\frac{y+2x}{2}\de_{\mu\nu ,\al\be}R^{\rh\la}
- \frac{y}{4}\,g_{\mu\nu}g_{\al\be}R^{\rho\la}
\,,
\label{D}
\eeq
\beq
(\hat{W})_{\mu\nu ,\al\be} & = &
\frac{3x}{2}
g_{\nu\be}R_{\mu\si\ta\th}R_{\al}\mbox{}^{\si\ta\th}
+\frac{x-y}{2}
R^\si\mbox{}_{\al\mu}\mbox{}^\ta R_{\nu\be\si\ta}
+\frac{5x+y}{2}
R^\si\mbox{}_{\al\mu}\mbox{}^\ta R_{\si\nu\be\ta}
\nonumber
\\
& + & \frac{3x+y}{2}\,
R^\si\mbox{}_\mu\mbox{}^\ta\mbox{}_\nu R_{\si\al\ta\be}
+ \frac{y-5x}{2}\,R_{\mu\ta}R^\ta\mbox{}_{\al\nu\be}
+ \frac{y+2x}{2}R_{\mu\al}R_{\nu\be}
\nonumber
\\
& + &
\Big( \frac{z}{2}R
- \frac{1}{4\ka^2}\Big) (R_{\mu\al\nu\be}
-\frac12\, g_{\mu\nu}R_{\al\be} + 3 g_{\nu\be}R_{\al\mu})
+ zR_{\mu\nu}R_{\al\be}
\nonumber
\\
& - & \frac{1}{4}
\Big[ xR_{\rh\la\si\ta}^2 + yR_{\rh\la}^2 + zR^2
- \frac{1}{\ka^2}(R-2\La )\Big]\,
\Big(\,\de_{\mu\nu\, ,\,\al\be}
-\frac12\,g_{\mu\nu}g_{\al\be}\,\Big)
\nonumber
\\
& + &
 \frac{3y}{2}g_{\nu\be}R_{\mu\ta}R^\ta\mbox{}_\al
-  xg_{\mu\nu}R_{\al\th\si\ta}R_\be\mbox{}^{\th\si\ta} -
yg_{\mu\nu}R_{\al\ta}R^\ta\mbox{}_\be\,. \label{W}
\eeq
In the
last two expressions we have used special condensed notations
which help presenting them in a relatively compact way. All
algebraic symmetries are implicit, including the symmetrization in
the couples of indices $\,(\al\be)\leftrightarrow(\mu\nu)$,
$\,(\al\leftrightarrow\be)\,$ and $\,(\mu\leftrightarrow\nu)\,$
and in the couple $\,(\rh\leftrightarrow\la)\,$ in the operator
$\hat{D}^{\rh\la}$. In order to obtain the complete formula
explicitly, one has to restore all these symmetries. For example,
\beq
R_{\mu\rh}R^\rh\mbox{}_{\al\nu\be}\to \frac{1}{2} \left(
R_{\mu\rh}R^\rh\mbox{}_{\al\nu\be} +
R_{\al\rh}R^\rh\mbox{}_{\mu\be\nu} \right) \nonumber
\eeq
restores
the $\,(\al\be)\leftrightarrow(\mu\nu)$ symmetry. Finally, let us
notice that formulas \ (\ref{K}), (\ref{D}), (\ref{W}) \
coincide with similar expressions of \cite{avrbar}
(see the last reference therein) in the particular case
$\,\,1/\rho\to 0$ \ after imposing the corresponding constraints
on $(x,y,z)$ in (\ref{x y z}).

The solution for the inverse matrix $\,\,\hat{K}^{-1}\,\,$ can be
easily found in the form
\beq
\hat{K}^{-1}\,=\,
(\hat{K}^{-1})_{\mu\nu ,}\mbox{}^{\,\th\om} =
\frac{4}{4x+y}\left(\de_{\mu\nu ,}\mbox{}^{\,\th\om}\, -\,\Om \;\;
g_{\mu\nu}\, g^{\th\om}\right)\,, \nonumber \eeq where \beq \Om =
\frac{y+4z}{4x-4z+n(y+4z)}\,. \nonumber
\eeq
Let us remark that in
dimensional regularization the divergences of the expressions
$\,\ln\,\Det \hat{\cal{H}}\,$ and $\,\ln\,\Det
(K^{-1}\,\hat{\cal{H}})\,$ are the same, because an extra factor
$\,\hat{K}^{-1}\,\,$ is a $c$-number operator.

By straightforward algebra, we obtain the minimal operator
in the standard form, useful for application of the
Schwinger-DeWitt technique \cite{frts82,bavi85}
\beq
\hat{H}\,=\,\hat{K}^{-1}\,\hat{\cal H} \,=\,
\hat{1}\Box ^2 + \hat{V}^{\rh\la}\na_\rh\na_\la +\hat{U}\,.
\label{main}
\eeq
It is important to notice that the new expressions
\beq
\hat{V}^{\rh\la}=\hat{K}^{-1}\hat{D}^{\rh\la}
\qquad  \mbox{and} \qquad
\hat{U}=\hat{K}^{-1}\hat{W}
\eeq
do not possess the symmetry
$\,(\al\be)\leftrightarrow(\mu\nu)$, while all other
symmetries are preserved. The expressions
for the two matrices $\hat{U}=(\hat{U})_{\mu\nu\, ,\,\al\be}$
and $\hat{V}^{\rh\la}=(\hat{V}^{\rh\la})_{\mu\nu\, ,\,\al\be}$
are very bulky and we settle them in Appendix A.

The algorithm for the one-loop divergences of a minimal fourth order
operator (\ref{bilinear}) is the following  \cite{frts82} \ \ (see
also \cite{bavi85} for an alternative, more systematic derivation):
\beq
\frac12 \ln \Det \hat{H}\Big|_{\rm div}
& = & -\,\frac{\mu^{n-4}}{(4\pi )^2(n-4)}\, \int d^nx\sqrt{g}\,
\tr \lim_{x^\prime \to x} a_2(x^\prime ,x)
\,=\,-\,\frac{\mu^{n-4}}{n-4}\,A_2\,,
\label{aa2}
\eeq
where
\beq
\lim_{x^\prime \to x}\,a_2(x^\prime\,,\,x)
& = & {\frac {\hat{1}}{90}}
\,R_{\rho\la\ka\om}^2-{\frac {\hat{1}}{90}}\,R_{\rho\la}^2+
\frac{\hat{1}}{36}\,R^2 \,+\,
\frac{1}{6}\hat{\cal{R}}_{\rho\la}\hat{\cal{R}}^{\rho\la}
-\hat{U}
\nonumber
\\
& - & \frac{1}{6}R_{\rh\la}\hat{V}^{\rh\la}
+ \frac{1}{12}R\hat{V}^\rh\mbox{}_\rh
+\frac{1}{48}\hat{V}^\rh\mbox{}_\rh\, \hat{V}^\la\mbox{}_\la +
\frac{1}{24}\hat{V}_{\rh\la}\hat{V}^{\rh\la}\,. \label{gr-formula}
\eeq
Here $\hat{\cal{R}}_{\rho\la}$ is the commutator of the
covariant derivatives acting in the tensor $\,h^{\al\be}\,$ space,
\beq
\hat{\cal{R}}_{\rho\la}\,=\,[\na_\rho\,,\,\na_\la ]\,.
\label{commute}
\eeq
The particular traces of the expression
(\ref{gr-formula}) are collected in the Appendix B. The result
must be summed up with the contributions of ghosts \ \ $\,\ln\Det
\hat{\cal{H}}_{gh}\,$ \ and the weight operator \ $\,\ln\Det
Y^{\al\be}\,$, according to the formula (\ref{e}). The
corresponding expressions are also given in Appendix B. Let us
present just the overall result for the divergent part of the
effective action, in terms of parameters $\,\la\,$, $\,\rho\,$
and $\,\xi\,$
\beq
A_2^t & = & -\mu^{n-4}\,\int
d^nx\sqrt{g}\,\left\{\, \be_1(n)\, E+ \be_2(n)\,
C^2+\be_3(n)\,R^2\, + \frac{\be_4(n)}{\ka^2}\,R
\right.
\nonumber \\
& + & \left.
\be_5(n)\,\frac{\La}{\ka^2}\,
+\,\frac{\be_6(n)}{\ka^4} \right\}\,.
\label{finaldiv}
\eeq
The
coefficients ($\be$-functions, as we shall see later on) are given
by the expressions
\beq
\be_i(n) =
\frac{1}{(4\pi)^2}\,\Big[\,\de^{(0)}_i + \frac{\de^{(1)}_i}{\rho}
+ \frac{\de^{(2)}_i}{\rho^2}\,\Big]\,, \qquad i=(1,2,3,4,5,6)\,.
\label{beta's} \eeq
For the sake of convenience let us now replace
\ \ $n = 4 - \ep$, such that all coefficients \
$\de^{(j)}_i(\xi\,,\la)$ \ should be expressed in terms of
$\,\ep$. The explicit form of the coefficients $\,\de^{(j)}_l$ \
corresponding to $\,\be$-functions for $\,\rho$, $\,\la\,$ and
$\,\xi\,$ are collected in Appendix C. Their general dependence on
the couplings has the following structure.
For $\be_1$ we have
\beq
\de_1^{(0)} =
\de_{1A}^{(0)} + \de_{1B}^{(0)}\,\,\frac{\la}{\xi}\, ,
\quad
\de_1^{(1)} =
\de_{1A}^{(1)}\,\xi + \de_{1B}^{(1)}\,\,\la\, ,
\quad \de_{1}^{(2)} =
\de_{1A}^{(2)}\,\xi^2
+ \de_{1B}^{(2)}\,\,\xi\la + \de_{1C}^{(2)}\,\,\la^2\,.
\label{final 1}
\eeq
For $\be_2$ we have the expressions
\beq
\de_2^{(0)} =
\de_{2A}^{(0)}\,\,\frac{\xi}{\la} + \de_{2B}^{(0)}
+ \de_{2C}^{(0)}\,\frac{\la}{\xi}\, ,
\quad
\de_2^{(1)} = \de_{2A}^{(1)}\,\xi
+\de_{2B}^{(1)}\,\,\la\, ,
\nonumber
\\
\de_2^{(2)} = \de_{2A}^{(2)}\,\xi^2
+\de_{2B}^{(2)}\,\xi\la
+ \de_{2C}^{(2)}\,\la^2\,.
\label{final 2}
\eeq
The coefficients of $\be_3$ can be written as
\beq
\de_3^{(0)} = \de_{3A}^{(0)}+\de_{3B}^{(0)}\,\frac{\la}{\xi}
+ \de_{3C}^{(0)}\,\,\frac{\la^2}{\xi^2}\, ,
\qquad
\de_3^{(1)} = \de_{3A}^{(1)}\,\xi
+ \de_{3B}^{(1)}\,\la
+ \de_{3C}^{(1)}\,\frac{\la^2}{\xi}\,,
\nonumber
\\
\de_3^{(2)} = \de_{3A}^{(2)}\,\xi^2
+ \de_{3B}^{(2)}\,\xi\la
+ \de_{3C}^{(2)}\,\la^2\, ,
\label{final 3}
\eeq
Furthermore, the coefficients $\de_4^{(j)}$ have the form
\beq
\de_4^{(0)} = \de_{4A}^{(0)}\,\xi
+ \de_{4B}^{(0)}\,\la
+ \de_{4C}^{(0)}\,\frac{\la^2}{\xi}\, ,
\qquad
\de_4^{(1)} = \de_{4A}^{(1)}\,\xi^2
+ \de_{4B}^{(1)}\,\,\la^2\, ,
\quad
\de_4^{(2)} = 0\,.
\eeq
Finally, the coefficients of functions
\ $\be_5$ \ and \ $\be_6$ \ have the form
\beq
\de_5^{(0)} =  \de_{5A}^{(0)}\,\xi
+ \de_{5B}^{(0)}\,\la \, ,
\qquad
\de_6^{(0)} =  \de_{6A}^{(0)}\,\xi^2
+ \de_{6B}^{(0)}\,\la^2\, ,
\qquad
\de_5^{(1,2)} = \de_6^{(1,2)} = 0\,.
\label{final 4}
\eeq
The coefficients \ $\de_{i\,A,B,C}^{(k)}$ \
depend only on $\,\ep\,$ and are listed in Appendix C.

After taking the limit $n\to 4$ in coefficients $\be_i(n)$, we
arrive at the counterterm, that is the negative of the $n\to 4$
coefficient for the pole term
\beq
\De S = - \Ga_{{\rm div}}^{(1)}
& = & \frac{\mu^{n-4}}{(4\pi)^2 (n-4)} \,\int d^nx\sqrt{g}
\,\left\{\,\frac{133}{20}\, C^2 -\frac{196}{45}\, E \right.
\nonumber
\\
& + &
\left.
\Big(\frac{10\,\la^2}{\xi^2} - \frac{5\,\lambda}{\xi}
+ \frac{5}{36}\Big)\,R^2\,
+ \,\Big(\frac{\xi}{12\,\la} - \frac{13}{6}
- \frac{10\,\la}{\xi}\Big)\,\frac{\la}{\ka^2}\,R
\right.
\nonumber
\\
& + &
\left.
  \Big(\, \frac{56}{3}
- \frac{2\,\xi}{9\,\la}\,\Big)\,\frac{\la\,\La}{\ka^2} +
\Big(\,\frac{\xi^2}{72} +
\frac{5\la^2}{2}\,\Big)\,\frac{1}{\ka^4}\,\right\}\,. \label{final
in 4}
\eeq
The last expression does agree with the well known result of
Avramidi and Barvinsky \cite{avrbar} \footnote{Including the
coefficient $\,(5/36)\,$, all coefficients agree with the previous
calculation of \cite{frts82}.}. This coincidence is remarkable in
several aspects. First of all, one can see that the effect of the
Gauss-Bonnet term is not relevant for one-loop
renormalization. This means that, in the framework of fourth
derivative quantum gravity, the hypothesis of Capper and Kimber
\cite{capkim} concerning the relevance of the topological term on
quantum level may be valid for finite corrections (e.g., this
is the case for the $\,4-\ep\,$ renormalization group)
or for sub-leading divergences at higher loops  but not for
the leading logarithms, including the one-loop divergences. Since
the theory under consideration involves all degrees of freedom of
the quantum metric, it is likely that the same situation holds in
{\it any} theory of quantum gravity. \ Second, the enormous
cancellation of the $\rho$-dependence (see Appendix C) in \
$\ep\to 0$ \ limit may be considered as a very strong test for
the result (\ref{final in 4}) and \cite{avrbar}. Let us
remark that Ref. \cite{avrbar} corrected the result of
previous pioneering calculations \cite{julton,frts82}, and
therefore an extra verification does not look unnecessary,
especially in view of existing applications \cite{applications}.
One has to notice that the difference between the results of
\cite{avrbar} and \cite{frts82} was only in $R^2$-term and
that the work \cite{avrbar} included an additional test
(derivation using $\ze$-regularization in a space of constant
curvature) for some special
combination of this and $C^2$-term. That test could be
inefficient only for the error proportional to a combination \
$R_{\rho\la}^2-\frac14 R^2$. Our calculation represents a very
strong test of this possibility and one may be certain that
the result (\ref{final in 4}) is a correct one. Third, our
calculations have been organized in such a way that we could
separate the result for the conformal case \ $\frac{1}{\xi} \to
0\,$ at every stage. Therefore, the coincidence of the final
result with the one of \cite{avrbar} is an additional confirmation
for our previous derivation of divergences in conformal
quantum gravity \cite{Weyl}, where we met a perfect agreement with
the result of \cite{antmot} and partial agreement with the one in
\cite{frts82}. Finally, the cancellation of a $\rho$-dependence in
the \ $\ep\to 0$ \ limit provides a particular but rather strong
test for the correctness of \ $\ep\neq 0$ \ coefficients
presented in (\ref{final 1})-(\ref{final 4}) and Appendix C. These
coefficients will be applied in the next section for investigating
the $\,4-\ep\,$ renormalization group equations.

\section{\large\bf  Renormalization group equations for higher
derivative terms}

Despite the $\rho$-dependence cancels out in the \ $\ep\to 0$ \
limit, the \ $4-\ep$ \ renormalization group equations do not
assume exactly the form of the known equations in four dimensions.
The reason is that \ $4-\ep$ \ renormalization group
$\be$-functions are sensitive to ${\cal O}(\ep)$-corrections
which depend on $\rho$. A standard derivation (see, e.g.
\cite{book}) yields the following renormalization group equations
for the parameters of the higher derivative sector
\beq
\frac{d\rho}{dt} & = & -\ep\,\rho + \be_1\,\rho^2
\nonumber
\\
\frac{d\la}{dt} & = & -\,\ep\,\la - 2\,\be_2\,\la^2
\nonumber
\\
\frac{d\xi}{dt} & = &  -\, \ep\,\xi - \be_3\,\xi^2\,, \label{RGE1}
\eeq
where $\mu$ is the renormalization parameter of dimensional
regularization, \ $dt = d\mu/\mu$ and the \ $\be$-functions are
given by formulas (\ref{beta's}). In the limit \ $\ep\to 0$, we
meet usual four dimensional renormalization group equations, which
are exactly the ones obtained earlier in \cite{avrbar}
\beq
(4\pi)^2\,\frac{d\rho}{dt} &=&
-\,\frac{196}{45}\,\rho^2\,,\qquad\qquad (4\pi)^2\,\frac{d\la}{dt}
= -\,\frac{133}{10}\,\la^2\,, \nonumber
\\
\nonumber
\\
(4\pi)^2\,\frac{d\xi}{dt}
&=& -\,10\,\la^2 + 5\,\la\,\xi - \frac{5}{36}\,\xi^2\,.
\label{betas 4}
\eeq

It is easy to see that the \ $n=4-\ep$ \ equations (\ref{RGE1})
are much more complicated than their \ $4$-dimensional
cousins (\ref{betas 4}). In order to see this, let us notice
that the first two equations (\ref{betas 4}) do not depend on
parameter $\xi$. These two equations can be easily
solved and their solutions replaced into the last
equation, which can be also solved analytically
\cite{frts82,avrbar}. For convenience we present the known
solution of all three equations and their analysis in Appendix D.
However, the equations (\ref{RGE1}) do not admit any simple
factorization and have to be solved simultaneously. Taking into
account their complexity, there is no hope to achieve solution in
the analytical form, and one has to apply numerical methods.

In fact, the situation with \ $4-\ep$ \
renormalization group equations is even more complex,
because one has to account for the arbitrariness coming
from the choice of a gauge-fixing condition. We have
already considered this aspect of theory in section 3.
Taking the gauge-fixing arbitrariness (\ref{total n})
into account, we arrive at
the complete form of  \ $4-\ep$ \ renormalization
group equations for the three parameters
\beq
\frac{d\rho}{dt} &=& -\ep\,\rho \,+\, \ep\,\rho\, f(\al_i)
+ \rho^2\,\be_1 \,,
\label{RGE1 rho}
\\
\nonumber
\\
\frac{d\la}{dt} &=& -\ep\,\la \,+\, \ep\,\la\, f(\al_i)
- 2\la^2\,\be_2 \,,
\label{RGE1 lambda}
\\
\nonumber
\\
\frac{d\xi}{dt} &=& -\ep\,\xi \,+\, \ep\,\xi\, f(\al_i)
- \xi^2\,\be_3 \,,
\label{RGE11}
\eeq
where the
$\be$-functions (\ref{beta's}) correspond to minimal
gauge-fixing condition. The remarkable feature of the
gauge-fixing dependent terms is that they are proportional
to the same quantity \ $f(\al_i)$. Furthermore, the
$\be$-functions (\ref{beta's}) are homogeneous functions
on the couplings. Taking all that into account,
our main strategy in the investigation of the system
(\ref{RGE1 rho}) -- (\ref{RGE11}) will be the
following. We construct two combinations $\th$ and $\om$
of the effective charges $(\la,\rho,\xi)$ such that the
renormalization group equations for these parameters are
free from gauge-fixing ambiguity. The equations for
the charges $\th$ and $\om$ will be explored with the
main purpose of establishing the UV stable fixed points.
After that we shall consider the equation for the
remaining effective charge, with the invariant
combinations $\th$ and $\om$ at the fixed point. In this
way we shall learn the asymptotic UV behaviour
corresponding to a given fixed point.

The most natural choice for independent effective charge is of
course $\la$, for it defines the interaction between gravitons.
Therefore we define the invariant charges as ratios between other
parameters and $\la$
\beq
\th = \frac{\la}{\rho} \,,\qquad \om =
-\,\frac{3\,\la}{\xi} \,.
\eeq
The coefficient in the second
expression provides correspondence with the notations of
\cite{frts82,avrbar}. It is straightforward to see that the
renormalization group equations for $\th$ and $\om$ are
independent on function  \ $f(\al_i)$ \ and have the
following universal form:
\beq
\frac{d\th}{d\tau} = -\,2\th\,\be_2 - \be_1
\,,\qquad\qquad \frac{d\om}{d\tau} =-\,2\om\,\be_2-3\be_3 \,.
\label{RGE2}
\eeq
Here $\tau(t)$ is a new parameter defined by
\beq
d\tau \,=\, \frac{\la(t)}{(4\pi)^2}\,\,dt \,,
\label{star}
\eeq
where \ $\la(t)$ \ is a solution of (\ref{RGE1 lambda}).

Equation (\ref{star}) can be viewed as a most important relation
of the whole $\,4-\ep$ approach. We know that \ $t\to \infty$ \
corresponds to the high energy limit $\,\mu\to\infty\,$ in the
$\,\overline{\rm MS}\,$ renormalization scheme. However, due to
the gauge-fixing dependence of the renormalization group
equation for $\,\la(t)$, it is not clear whether this high energy
limit corresponds to a certain limit in the new variable $\,\tau$.
Perhaps even more important aspect of the same problem is the
possible arbitrariness in the asymptotic behaviour of the
coupling $\,\la$.
This arbitrariness may put under question the asymptotic
freedom of the theory, spoiling the physical sense
of the renormalization group. We shall consider the asymptotic
behaviour of $\,\la\,$ later on and for a while, when looking
for the new fixed points for $\om$ and $\th$, will
identify the limit $\,\tau\to\infty\,$ as the UV one.

\vskip 1mm
\begin{center}
\begin{tabular}{|c||c|c||c|}
\hline
Fixed Point & $\th$ & $\om$ & Stability   \\
\hline\hline
1  & 0.33516  & -5.38892  &  Saddle \\
\hline
2  & 4.61183  &  -1.60198  &  UV-Unstable \\
\hline
3  &  -4.31710 & -1.47066 &  UV-Unstable  \\
\hline
4  &  -4.44192  &  -0.15162 &  Saddle \\
\hline
5  & 4.80565  & -0.06229  &  Saddle  \\
\hline
6  & 0.33782 & -0.00283  &  UV-Stable \\
\hline
7  & -3.94162  & 0.03123 &  UV-Unstable \\
\hline
8  & -4.11072  & 0.07230  &  Saddle \\
\hline\hline
\end{tabular}
\end{center}
\begin{quotation}
{\bf Table 1.} \ The list of the fixed points and
their stability with respect to small perturbations
for the case $\ep = 0.1$.
\end{quotation}
\vskip 1mm

According to the outline described above, consider equations
(\ref{RGE2}). For any given value of $\,\ep$, this is a system
of two nonlinear and rather complicated equations. There are no
chances to obtain the general solution analytically, so the
unique visible possibility is to fix the value of \ $\ep$,
according to the standard practice \cite{Wilson73}. We shall
try different values of \ $\ep$, starting from \ $\ep=0.1$.
In this case the equations for the fixed points
\beq
2\th\,\be_2 + \be_1 = 0
\,,\qquad
2\om\,\be_2 + 3\be_3 = 0\, ,
\label{fixed 0.1}
\eeq
have eight distinct solutions collected in the Table 1.
We investigated their stability under small perturbations
of both charges \ $\th_k \to \th_k + \de\th_k$,\
$\om_k \to \om_k + \de\om_k$, where the values \ $\th_k$,
\ $\om_k$, \ $k=1,2,..,8$ \ correspond to different fixed
points.
It turns out that for \ $\ep=0.1$ \ there are all kinds
of the fixed points: UV stable, UV-unstable in all directions
and saddle points.

In order to compare with the standard \ $\ep = 0$ \ case (see
Appendix D), we present the corresponding fixed points in the
Table 2. In this case there are only two fixed points, what means
that six new fixed points have emerged due to the procedure
$\,\ep = 0.1$.
The renormalization group trajectories which result from
numerical investigation are shown at Figure 1 for the simplest
$\,\ep = 0$ case. Each arrow indicates the slope of integral
curves, $\,d\th/d\om\,$ at a given point of the phase diagram.

\vskip 1mm
\begin{center}
\begin{tabular}{|c||c|c||c|}
\hline
Fixed Point & $\th$ & $\om$ & Stability   \\
\hline\hline
1  & 0.32749  &  -0.02286 &  UV-Stable \\
\hline
2  & 0.32749  &  -5.4671  &  Saddle \\
\hline\hline
\end{tabular}
\end{center}
\begin{quotation}
{\bf Table 2.} \ Fixed points for $\th$ and $\om$ and
their stability for the simple case\ $\ep = 0$
\ The equations admit also complete analytical
investigation presented in Appendix D.
\end{quotation}
\vskip 1mm

\begin{figure}
$$
\psfig{file=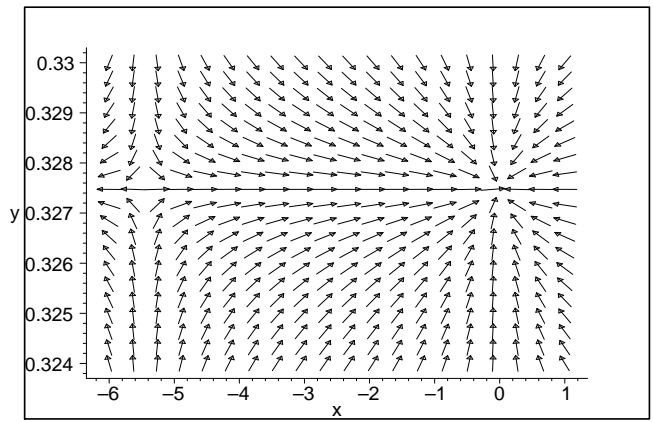,width=8.5cm,height=5.5cm}
$$
\vspace*{8pt} \caption{
Diagram for $\ep = 0$ (four dimensions). Here $y$ stands for
$\th$ and $x$ for $\om$. }
\end{figure}

For $\ep = 0.1$, the renormalization group flow is illustrated at
Figure 2. Notice that the points 1 and 6 are shown in detail in
the right diagram, which is very similar to the diagram at Figure
1. Indeed the trajectories in a limited region containing points 1
and 6 are practically the same trajectories carried out for $\ep =
0$. Thus, the procedure $\ep = 0.1$ keeps the dynamics unchanged
in some limited region of the plane\footnote{The meaning of this
statement is somewhat restrictive: of course one has to consider
only trajectories confined in that region.}, producing just a
small displacement of the fixed points (in this sense, points 1
and 6 are direct descendants from the standard 4D-fixed points).
However, in large scale, dynamics is totally modified by the
appearance of six extra fixed points.

The situation for other small positive values of $\ep$ is
qualitatively similar to the one in the $\ep = 0.1$ case. For
example, for $\ep = 0.01$ we meet 10 fixed points, but only
one of them is UV-stable. We will not present the details
of this case here.

The results for the negative $\ep$ are quite different.
The result of numerical integration for the cases
$\ep = -0.1$ and $\ep = -0.01$ are
shown at Figure 3. The general structure is similar to the
solution in four dimensions, and no extra fixed point emerges
by consequence of $\ep\neq 0$, only small displacements of these
points take place.

Now we are in a position to explore the issue of asymptotic
freedom in $4-\ep$ framework. For this end we need to
investigate the behavior of the effective charge \ $\la$ \ in the
UV limit \ $t\to\infty$. Also, this defines the physical sense of
parameter $\tau(t)$ and its possible relation to the change of
a physical energy scale. Hence, this consideration should clarify
the physical sense of equations (\ref{RGE2}) and their main
characteristics, that are the corresponding fixed points.

\begin{figure}
$$
\begin{array}{l}
\psfig{file=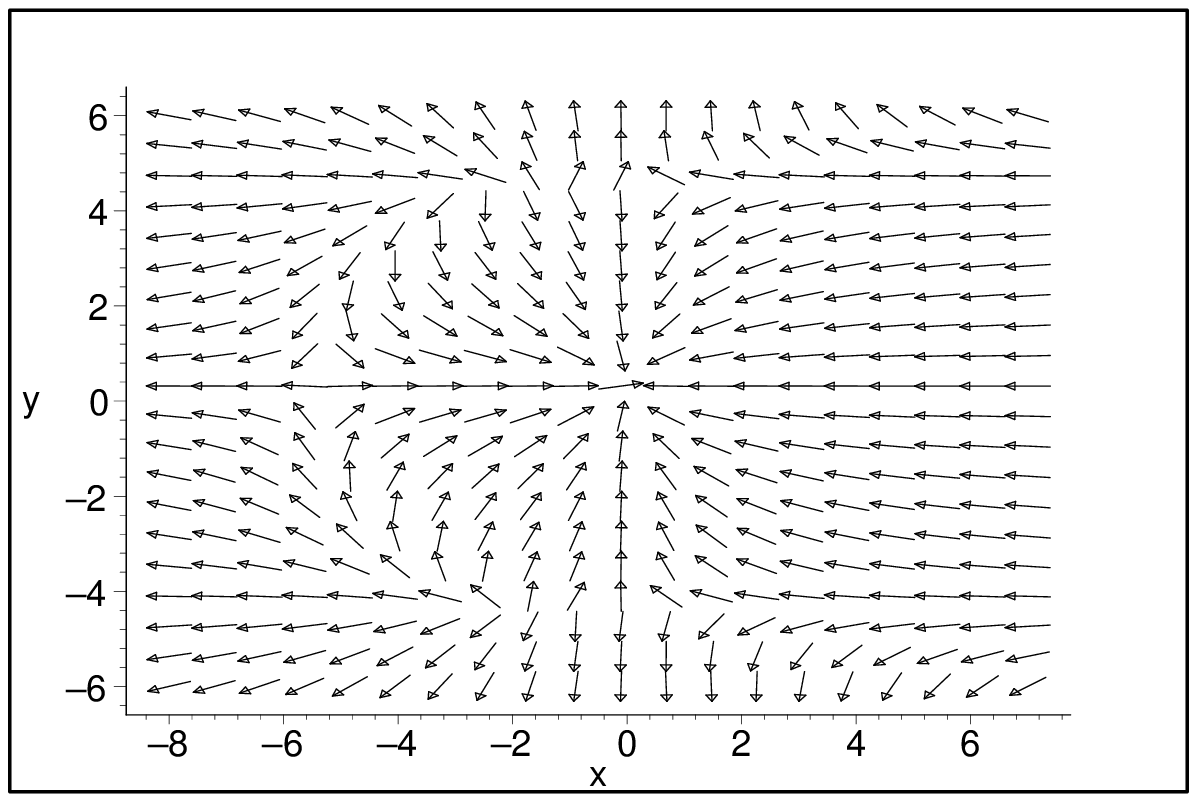,width=7.8cm,height=5.5cm}\;\;
\psfig{file=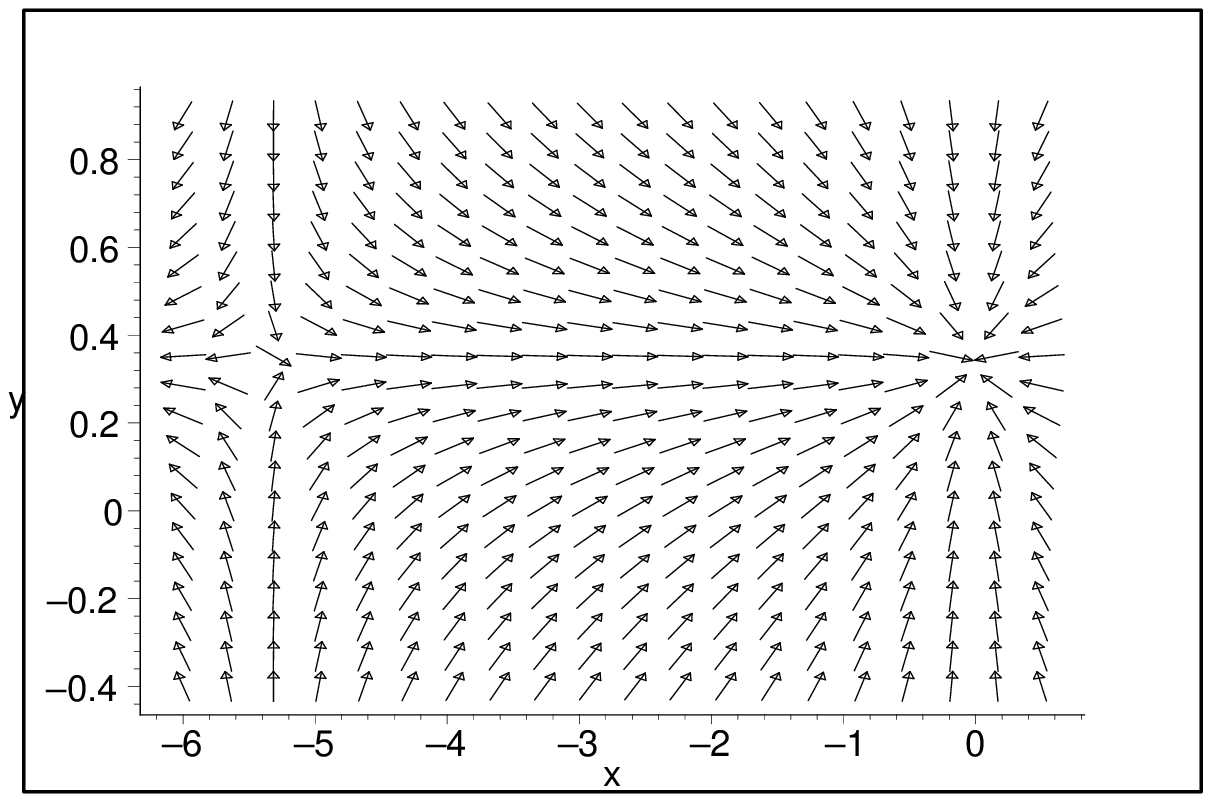,width=7.8cm,height=5.5cm}
\end{array}
$$
\vspace*{8pt} \caption{Numerical integration for $\ep = 0.1$ in
the whole plane ($\om$, $\th$) in the left diagram. The right
diagram shows a more detailed description of the trajectories
between point
1 (saddle) and point 6, which is UV-stable.} 
\end{figure}

Starting from the equation for $\,\la(t)\,$ in (\ref{RGE1}),
one can find the analytical form of $\la(t)$ in the vicinity
of a fixed point \ $(\om_0,\th_0)$. For this end,
the expression $\be_2$ must be rewritten in terms
of $\om$ and $\th$, which should be further replaced by
\ $\om_0$ \ and \ $\th_0$. After performing this,
independent on the values of
\ $\om_0$ \ and \ $\th_0$ \ we arrive at the equation
\beq
\frac{d\la}{dt} = a\la - b^2\la^2\,,
\label{eqLa}
\eeq
where
\beq
a = -\ep + \ep\, f(\al_i)
\qquad {\rm and} \qquad
b^2 = 2\,\be_2(\om_0\,,\th_0)\,.
\label{a b}
\eeq
Let us remark that the parameter $b$ in the last equation
depends only on the values $(\om_0\,,\th_0)$, while the
parameter $a$ depends also on the choice of a gauge-fixing
condition and therefore can be made arbitrary.

The solution of (\ref{eqLa}) is straightforward
\beq
\la(t) = \frac{a\la_0\,e^{at}}{b\la_0\,(e^{at}-1)+ a}\,,
\qquad \la_0 = \la(0)\,.
\label{lambda}
\eeq
Starting from this relation, we
integrate (\ref{star}) and arrive at the explicit form of $\tau$
\beq
\tau(t) = \frac{1}{b}\,\ln \, \left[a +
b\la_0\,(e^{at}-1)\right] + C\,,
\eeq
where $\,C\,$ is irrelevant integration constant.

\begin{figure}
$$
\begin{array}{l}
\psfig{file=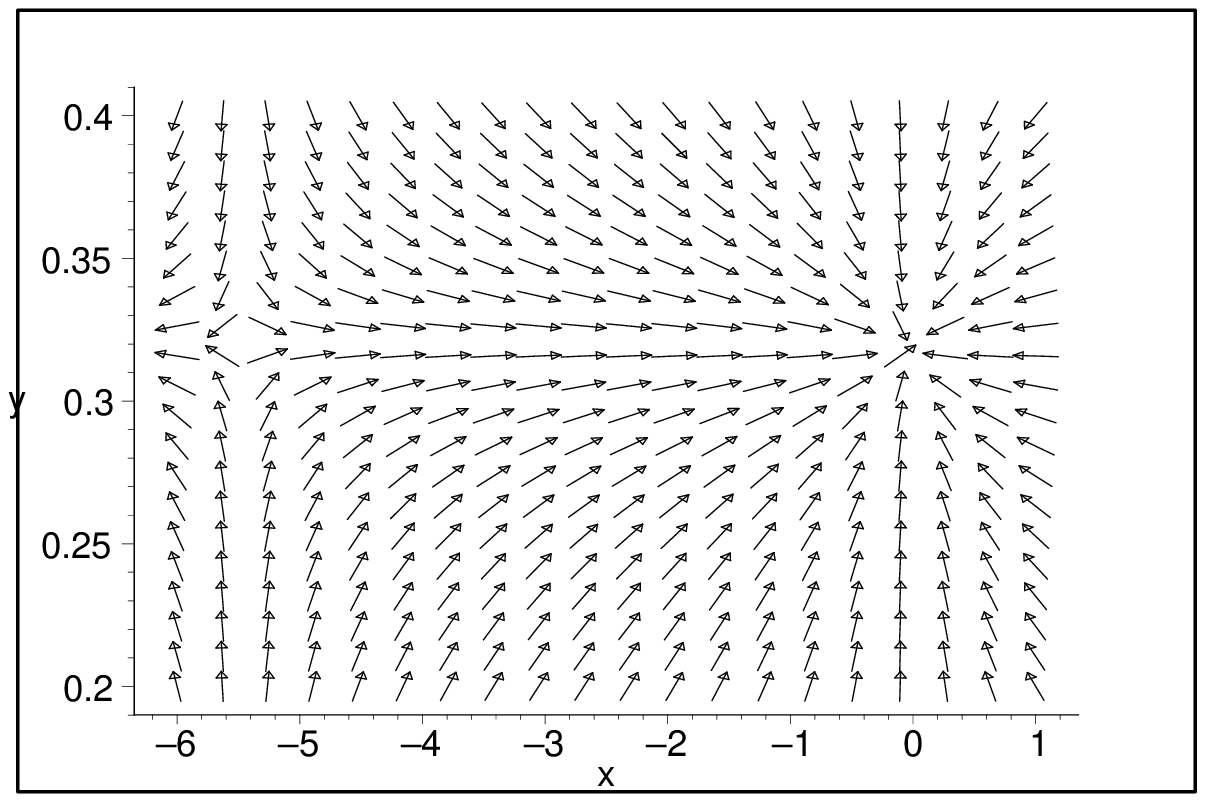,width=7.8cm,height=5.5cm}\;\;
\psfig{file=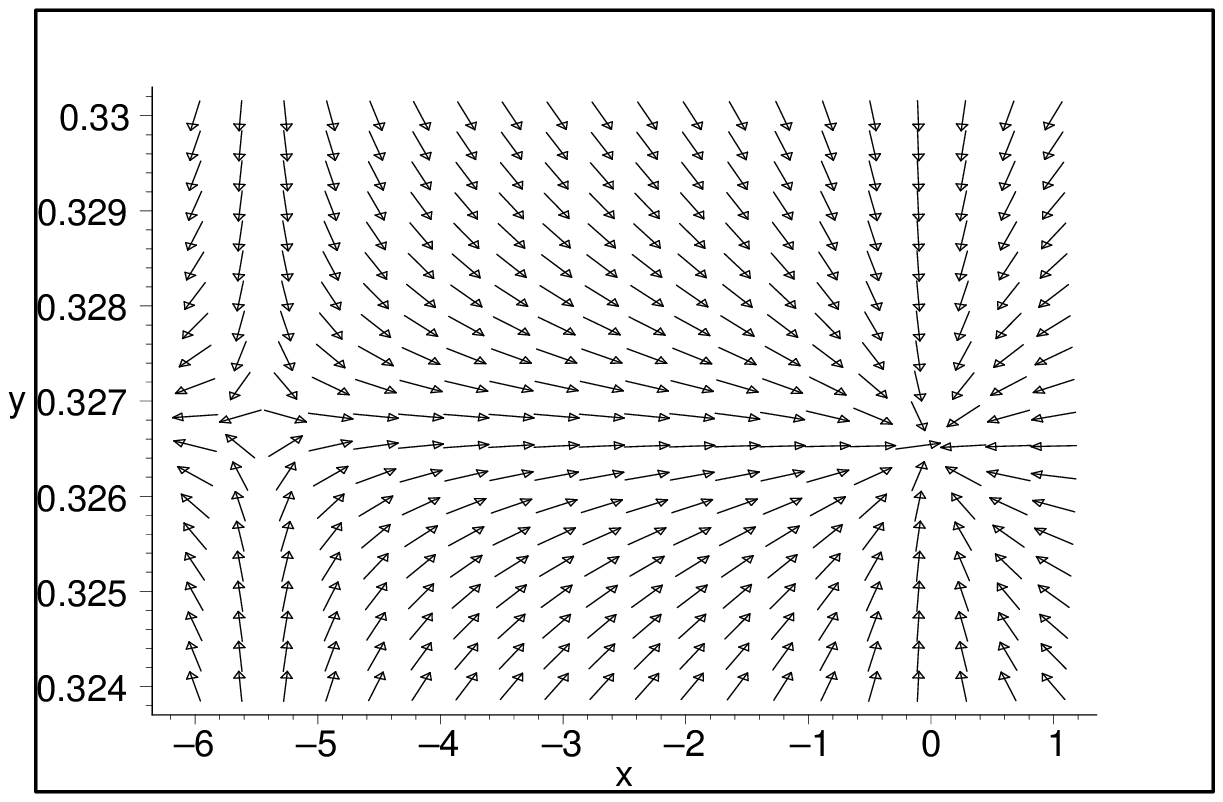,width=7.8cm,height=5.5cm}
\end{array}
$$
\vspace*{8pt} \caption{Numerical integration for $\ep = -0.1$
(left) and $\ep = -0.01$ (right), where the two fixed points in
both cases are easily recognizable.}
\end{figure}

It is easy to see that, formally, the asymptotic freedom in the
theory depends on the sign of the quantity $\,a\,$ in equation
(\ref{a b}). In case \ $a>0$ \ we have $\,\la(t)\to 0\,$ in the UV
$\,t\to \infty$, and also $\,\tau\to\infty$, so everything is
consistent with the $\,\ep=0\,$ case. On the opposite, for
$\,a \leq 0\,$ we meet the asymptotic behaviour
$\,\la(t)\to \la_0 \neq 0\,$
in the UV limit $\,t\to \infty$, also $\,\tau\to \tau_0 \neq \infty$.

It is important that the explicit form of the function
\ $f(\al_i)$ \ derived in \cite{shja}
\beq
f(\al_i=\alpha,\gamma)\,=\,\frac32\,
\left\{\, \frac{\la}{2}\,\ln \Big[
\frac{9\,\ga\,\la^{5/2}}{2\,\al^{3/2}\,(\al-6\la\ga)}\Big]
\,\,+ \,\,\al\,\,-\,\,\la \right\}
\label{explicit}
\eeq
does not produce any restrictions for the value of
\ $f(\al_i)$, which can be modified arbitrarily by
choosing one or another gauge-fixing condition.
According to equation (\ref{a b}), this means that
one can also change the sign of the quantity $a$,
and thus the asymptotic freedom in the  \ $4-\ep$ \
theory is not a physical phenomenon, but an
artificial occurrence depending on the choice of the
gauge-fixing condition.

Is it true that $\,4-\ep\,$ renormalization group does not
have physical sense for the case of higher derivative quantum
gravity? In fact, this is a problem of interpretation. Let
us remind that the original $\,4-\ep\,$ approach \cite{Wilson73}
treats $\,\ep\,$ as a small parameter of the non-perturbative
expansion. If we take this position and consider all terms
proportional to $\,\ep\,$ as small {\sl by definition}, then
the asymptotic behavior for \ $\la$ \ is close to \ $n=4$ \
renormalization group equation (\ref{a f 1}).
In this case the status of the new fixed points becomes
clear and we conclude that the stable
fixed points in all cases are similar to the one in \ $n=4$ \
renormalization group. The same concerns also the renormalization
group flows in the vicinity of the stable fixed points. At the
same time, the existence of the numerous new unstable and saddle
fixed points for \ $\ep>0$ \ indicates the possibility of a rich
non-perturbative structure of the theory.

\section{Renormalization Group in the Einstein-Hilbert sector}

Consider the renormalization group equations in the
Einstein-Hilbert sector of the theory, including the cosmological
constant. The renormalization group equations for the parameters
in the low energy sector of theory are defined by the expressions
for divergences (\ref{finaldiv}), but must also account for
the gauge-fixing dependence (\ref{total n}). The form of
these equations is as follows:
\beq
\frac{d\ka^2}{dt} = (n -
4)\,\ka^2 + \ka^2\beta_4 + \frac{n-2}{2}\ka^2 f(\al_i)
\label{Newton}
\eeq
for the Newtonian constant and \beq
\frac{d\La}{dt} = \La\be_4 + \frac12\,\La\be_5 +
\frac{\be_6}{2\ka^2} - \La\,f(\al_i)\, \label{C C}
\eeq
for
the cosmological constant.

The presence of the function $\,f(\al_i)\,$ in both equations
indicates a strong gauge-fixing dependence. Indeed, there is
nothing new in this dependence, as already known from
\cite{frts82} (see also \cite{shja}). The standard solution of this
problem is to look for the gauge-fixing invariant combination of
the two $\be$-functions. It is assumed that the corresponding
effective charge is an essential coupling constant
\cite{Weinberg79}. Indeed, in the
$n=4$ case such combination is the dimensionless
product \ $\ka^2\La$ \ of the two effective charges \ $\ka$ \ and
\ $\La$. The question is whether the essential effective charge
can be found for the case of $\,4-\ep\,$ renormalization group
equation.

By direct computation one can easily obtain the
unique combination
$$
\ga=\ka^2\La^N \,, \qquad N = \frac{n-2}{2}
$$
of \ $\ka$ \ and \ $\La$ \ which is explicitly independent on
gauge-fixing parameters. As one could expect, this is exactly the
combination of \ $\ka$ \ and \ $\La$ \ which is dimensionless for
any value of \ $n$. Of course, for \ $n=4$ \ we come back to the
standard combination of parameters \ $\ga(4)=\ka^2\La$.

Using (\ref{Newton}) and (\ref{C C}) we arrive at the
renormalization group equation for \ $\ga$
\beq
\frac{d\ga}{dt}
\,=\, \ga\,\left[\, n-4 + \frac{1}{2}\left( 2(N+1)\be_4 +
N\be_5 + \frac{N}{\ka^2\La}\be_6\right)\right]\,. \label{gamma}
\eeq
The last equation looks like the one for the essential
coupling, but in fact there is still a difficulty in its
interpretation. The problem is that the expressions $\be_{4,5,6}$
can be written in terms of essential couplings, $\om$ and $\th$,
but there will be some extra factors of $\la(t)$ in (\ref{gamma})
such that it can not be absorbed in a redefinition of $dt$.
Besides, the gauge dependent quantity $\ka^2\La$ (which is
dimensionless only in four dimensions) also appears in the
coefficient of $\be_6$. The consequence is that, without
imposing the smallness of $\,\ep$, the
$(4-\ep)$-renormalization group for $\ka^2\La^N$ is gauge
dependent, exactly as in the case of the higher derivative
couplings. On the other hand, if we remember that all our
consideration is called to model the non-perturbative
$\ep$-expansion and define $\,\ep\,$ as a small parameter,
the renormalization group flow near the UV-stable fixed
point $\,\om_0,\,\th_0\,$ will not have qualitative
difference with the $\,\ep=0\,$ case investigated in
\cite{frts82,avrbar}. The main feature of the UV asymptotic
behaviour of $\,\ga\,$ is strong dependence on the initial
data for the couplings $\,\xi\,$ and $\,\ga\,$.

The most interesting aspect of the renormalization group
equation for the cosmological constant would be to study the
low-energy limit. This investigation could provide better
understanding of the possible role of quantum gravity for
the cosmological constant problem \cite{Polyakov,tsamis},
including the remnant quantum effects of the particles with
the mass of the Planck order of magnitude \cite{cosm} and
also for establishing the assumed universality property of
quantum gravity \cite{don}. However, the framework presented
above is not appropriate for this purpose, because in the
low-energy region we expect to meet a decoupling of the
massive (ghost) mode (\ref{ghost}) and the renormalization
group equation should essentially modify. In order to observe
decoupling one has to perform calculations and find
$\be$-functions in the physical mass-dependent
renormalization scheme. Similar program has been realized
recently for the massive matter fields on curved background
\cite{apco}. The practical calculations in higher derivative
quantum gravity, despite they are looking rather difficult
from the technical point of view, represent a serious
challenge for the future work.

\section{Conclusions}

We performed  a complicated calculation of the counterterms in the
fourth derivative quantum gravity. It is the first time that the
quantum effects of the Gauss-Bonnet term have been taken into
account. In the $n=4-\epsilon$ case the effects of this term are
shown to be non-trivial, in accordance with the prediction by
Capper and Kimber \cite{capkim} concerning an important role of
the topological term in quantum gravity. At the same time, for
$n=4$ the effects of the topological term cancel, in this limit
we recover the known result \cite{frts82,avrbar}. This
coincidence represents very efficient verification of
previous derivations of the $n=4$ renormalization group equations.

The renormalization group equations for the parameters of theory
in the $n=4-\epsilon$ are essentially more complicated than in
the $n=4$ case, in
particular, the equations for different couplings do not separate.
Moreover these equations manifest much stronger gauge-fixing
dependence than in the standard
$n=4$ case. We separated the universal corner of theory, which is
characterized by a number of new fixed points for $\,\ep>0$.
These fixed points may be related to the rich non-perturbative
structure of the theory in the $\,\ep$-expansion.
Out of the small-$\ep\,$ approximation the equation for the most
important
coupling $\la$ (the parameter of the loop expansion of the
theory), manifests a strong gauge-fixing dependence which does
not take place for the usual $n=4$ renormalization group. From
a purely formal
point of view, the physical interpretation of the $n=4-\epsilon$
renormalization group looks unclear in this case.
However the situation changes if we
treat $\,\ep\,$ as a small parameter. Then the theory remains
asymptotically free for $\,\ep \neq 0\,$ and moreover, as we
already mentioned, has a number of new fixed points for $\ep >0$.

\vskip 10mm

\noindent {\large\bf Acknowledgments.} One of the authors (G.B.P.)
is grateful to Departamento de F\'{\i}sica at the Universidade
Federal de Juiz de Fora for kind hospitality. The work of the
authors has been supported by the fellowship from FAPEMIG (G.B.P.)
and the research grant from CNPq (I.Sh.). We are grateful to
Roberto Percacci for indicating several misprints in the previous
version of this preprint.

\vskip 10mm

\section*{\large\bf  Appendix A}

In this Appendix we collect bulky
elements of the operator (\ref{main}). The expression
for $\hat{U}=(\hat{U})_{\mu\nu\, ,\,\al\be}$
has the following form:
\beq
(\hat{U})_{\mu\nu\, ,\,\al\be}
& = & \frac{4}{y+4x}\, \left\{\frac{3x}{2}\,
g_{\nu\be}R_{\mu\rh\la\si}R_\al\mbox{}^{\rh\la\si}
+\frac{5x+y}{2}R^\la\mbox{}_{\al\mu}\mbox{}^\rh R_{\la\nu\be\rh}
+\frac{x-y}{2}R^\rh\mbox{}_{\al\mu}\mbox{}^\la R_{\nu\be\rh\la}
\right.
\nonumber \\
& + & \left.
\frac{y-5x}{4}\,\Big( R_{\mu\si}R^\si\mbox{}_{\al\nu\be}+
R_{\al\si}R^\si\mbox{}_{\mu\be\nu}\Big)
+ \frac{y+2x}{2}R_{\mu\al}R_{\nu\be}
- \frac{x}{2}g_{\al\be}R_{\mu\th\si\ta}R_\nu\mbox{}^{\th\si\ta}
\right.
\nonumber \\
& - & \left. \frac{1}{4}\,S^2
\,\left( \de_{\mu\nu\, ,\,\al\be}-\Om_1\,g_{\mu\nu}g_{\al\be} \right)
+ \left(\frac{3zR}{2}-
\frac{3}{4\ka^2}\right)g_{\nu\be}R_{\al\mu}
+ \frac{3y}{2}g_{\nu\be}R^\la\mbox{}_\mu R_{\al\la}
\right.
\nonumber \\
& + & \left.
\frac{x+3y}{4}\,g_{\mu\nu}\,R^{\si\ta}R_{\al\si\be\ta}+
\left(\frac{zR}{2}-
\frac{1}{4\ka^2}\right)R_{\mu\al\nu\be}
+ \frac{3x+y}{2}\,
R^\si\mbox{}_{\mu}\mbox{}^\ta\mbox{}_\nu R_{\ta\al\si\be}
\right.
\nonumber \\
& - & \left.
\frac{y}{2}g_{\al\be}R_\mu\mbox{}^\ta R_{\nu\ta}-
x\,\Om_1\,g_{\mu\nu}R_{\al\th\si\ta}R_\be\mbox{}^{\th\si\ta}-
\Om_2\,g_{\mu\nu}R_\al\mbox{}^\ta R_{\be\ta}+
zR_{\mu\nu}R_{\al\be}
\right.
\nonumber \\
& + & \left.
\left( \frac{1}{4\ka^2} -\frac{zR}{2}\right)g_{\al\be}R_{\mu\nu}+
\left( \frac{1}{\ka^2}\,\Om_3
- z\,R\,\Om_1\,\right) g_{\mu\nu} R_{\al\be}
\right.
\nonumber \\
& - & \left. \frac{1}{4\ka^2}\,\Om \,
\left(4\La - R\right)\,g_{\mu\nu}g_{\al\be}
\right\}\,,
\label{U}
\eeq
where
\beq
S^2=xR^2_{\rh\la\si\ta}+yR^2_{\rh\si}+zR^2-\frac{1}{\ka^2}
\left(R-2\La\right)\, ,
\eeq
and the coefficients $\Om_{1,2,3}$ are given by
\beq
\Om_1 & = & \frac{2\,x + 3\,y + 10\,z}
  {\Si }\, ,
  \;\;\;\;
\Om_2 = \frac{ 7\,x\,y + 9\,y^2 - 4\,x\,z + 28\,y\,z }
  {4\,\Si }\, ,
\nonumber \\
\Om_3 & = & \frac{x + y + 3z}{\Si}
\, , \;\;\;\;\;
{\rm with}\;\;\;
\Si =  4\,x - 4\,z + n\,\left( y + 4\,z \right)\,.
\eeq

For the matrix \ $\hat{V}^{\rh\la}
=(\hat{V}^{\rh\la})_{\mu\nu\, ,\,\al\be}$ \ we have the
expression
\beq
\hat{V}^{\rh\la} &=& \frac{4}{y+4x}\sum^{20}_{i=1}
\, b_i\, {\bf k}_i\,,
\label{V}
\eeq
where the following condensed notations have been used:
\beq
{\bf k}_1 &=&  g_{\nu\be}\,g^{\rh\la}\,R_{\mu\al}
\,,\qquad
{\bf k}_2 = \de_{\mu\nu\, ,\,\al\be}\,g^{\rh\la}
\,,\qquad
{\bf k}_3 = g^{\rh\la}\,R_{\mu\al\nu\be}\,,
\nonumber
\\
{\bf k}_4 &=& \de_{\nu\be ,}\mbox{}^{\rh\la}\,R_{\mu\al}
\,,\qquad\,\,\,
{\bf k}_5 = \de_{\nu\be ,}\mbox{}^{\rh\la}\,g_{\mu\al}
\,,\qquad
{\bf k}_6 = \de_{\mu\nu ,\,\al\be}\,R^{\rh\la}\,,
\nonumber
\\
{\bf k}_7 &=& \frac12(\,\de_\nu^{(\rh} \,R^{\la )}\mbox{}_{\al\be\mu}
+\de_\be^{(\rh} \,R^{\la )}\mbox{}_{\mu\nu\al}\, )
\,,\qquad\qquad
{\bf k}_8  = g_{\nu\be}\,\de_{(\mu}^{(\rh} \,R^{\la )}\mbox{}_{\al )}\,,
\nonumber
\\
{\bf k}_9  &=&  g_{\nu\be}\,R_{(\al}\mbox{}^{\rh\la}\mbox{}_{\mu )}
\,,\qquad
{\bf k}_{10} = \frac12\,(\,\de_{\al\be ,}\mbox{}^{\rh\la}\,R_{\mu\nu}
+\de_{\mu\nu ,}\mbox{}^{\rh\la}\,R_{\al\be}\, )\,,
\nonumber
\\
{\bf k}_{11} &=& g_{\mu\nu}R_\al\mbox{}^{\rh\la}\mbox{}_\be
\,,\qquad\,\,\,\,
{\bf k}_{12} = g_{\al\be}R_\mu\mbox{}^{\rh\la}\mbox{}_\nu
\,,\qquad
{\bf k}_{13} = g_{\mu\nu}\,g^{\rh\la}R_{\al\be}\, ,
\nonumber
\\
{\bf k}_{14} &=& g_{\al\be}\,g^{\rh\la}R_{\mu\nu}
\,,\qquad
{\bf k}_{15} = g_{\mu\nu}\,\de^\la_\al\,R^\rh_\be
\,,\qquad
{\bf k}_{16} = g_{\al\be}\,\de^\la_\mu\,R^\rh_\nu\,,
\nonumber
\\
{\bf k}_{17} &=& g_{\mu\nu}\,\de_{\al\be ,}\mbox{}^{\rh\la}
\,,\qquad\,\,\,
{\bf k}_{18} = g_{\al\be}\,\de_{\mu\nu ,}\mbox{}^{\rh\la}
\,,\qquad
{\bf k}_{19} = g_{\mu\nu}\,g_{\al\be}\,g^{\rh\la}\,,
\nonumber
\\
{\bf k}_{20} &=& g_{\mu\nu}\,g_{\al\be}\,R^{\rh\la}\,
\eeq
and
\beq
b_1 &=& -2x
\,,\qquad
b_2 = \frac{z R}{2}-\frac{1}{4\ka^2}
\,,\qquad
b_3 = 3x+y
\,, \qquad
b_4 = 2x
\nonumber
\\
b_5 &=& \frac{1}{2\ka^2} - z R
\,,\qquad
b_6 = x+\frac{y}{2}
\,,\qquad
b_7 = -4x
\,,\qquad
b_8 = -4x-2y  \,,
\nonumber
\\
b_9 &=& -2x
\,,\qquad
b_{10} = -2z
\, ,\qquad
b_{11} = 4x\,\Om_3
\,,\qquad
b_{12} = x
\, ,
\nonumber
\\
b_{13} &=& -y\,\Om_3
\,,\qquad
b_{14} = z
\,,\qquad
b_{15} = 2y\,\Om_3
\, ,\qquad
b_{16} = \frac{y}{2}
\,,
\nonumber
\\
b_{17} &=& 2zR\,\Om_3- \frac{\Om_1 -2\,\Om }{2\,\ka^2} \,,\qquad
b_{18}  = \frac{z R}{2}-\frac{1}{4\ka^2} \,,\qquad b_{19}  =
-b_{17} \,, \nonumber
\\
b_{20} &=& -y\,\Om_3\, .
\label{b}
\eeq
The last expressions have a non-symmetric form, and the rule
of symmetrization is the one described in the section 4.

\section*{\large\bf  Appendix B.
Particular results for necessary traces}

In this Appendix we collect some intermediate formulas, which
are necessary for the computation of the one-loop contributions.
The calculations has been verified by using the
package MathTensor \cite{MathTensor} driven by the software
Mathematica \cite{M4}.

The trace of the operator \ $\hat{U}$ \ is given by
\beq
\tr \hat{U} = \de^{\mu\nu , \al\be}\,U_{\mu\nu , \al\be}
= A_1\,R_{\mu\nu\al\be}^2 + A_2\,R_{\mu\nu}^2 +
A_3\,R^2 + A_4\,\frac{R}{\ka^2} + A_5\,\frac{\La}{\ka^2}\, ,
\eeq
where
$$
A_k = \frac{p_{k1} + p_{k2}\, n + p_{k3}\,n^2
+ p_{k4}\,n^3}{2\,\left( 4\,x + y \right)
\,\left( 4\,x - 4\,z + ny + 4nz \right)}\,,\qquad k=1,2,3,4,5
$$
and
\beq
p_{11} &=& 8(8x^2 - 20xz - 3yz)
\,, \qquad\quad
p_{12} = 2\left( 12x^2 + 13xy + 3y^2 + 40xz
+ 12yz \right) \, ,
\nonumber
\\
p_{13} &=& - x\,\left( 4\,x - 5\,y - 24\,z \right)
\,,
\qquad\qquad\quad\quad
p_{14} = - x\,\left( y + 4\,z \right) \, ,
\nonumber
\\
p_{21} &=&
8 (12x^2 - 3y^2 - 8 xz - 12yz - 4z^2)
,\,\,\,\,
p_{22} = 2( 24xy + 5y^2 + 48xz + 12yz + 16z^2),
\nonumber
\\
p_{23} &=& - y\,\left( 4\,x - 5\,y - 24\,z \right)
\,, \qquad\qquad\quad\quad
p_{24} = - y\,\left( y + 4\,z \right)\,,
\nonumber
\\
p_{31} &=& 8\left( 2x^2 + xy - 4xz - 4yz - 10z^2 \right)
,\quad
p_{32} = 2\left( 2xy + y^2 + 20xz + 7yz \right),
\nonumber
\\
p_{33} &=& - z\,\left( 4\,x - 5\,y - 24\,z \right)
,\qquad\qquad\qquad \,
p_{34} = - z\,\left( y + 4\,z \right)\,,
\nonumber
\\
p_{41} &=& 8 (x  + y + 3z)
,\,\,\,
p_{42} = - 4(y+z+3x)
,\,\,\,
p_{43} = 4x  - 2y - 12z
,\,\,\,
p_{44} = y + 4z  ,
\nonumber
\\
p_{51} &=& 0
\,,\qquad\,
p_{52} = 4\left( y + 4z \right)
,\qquad\,\,
p_{53} = - 2\left( 4x + y \right)
,\qquad\,\,
p_{54} = - 2\left(y + 4z \right)\,.
\nonumber
\eeq

The expression for \ \ $\tr (\hat{V}_\ga^\ga R)$ \ can be
presented as
\beq
\tr (\hat{V}_\ga^\ga R)\,=\,B_1\,R^2\,+\,\frac{B_2}{\ka^2}\,R\, ,
\eeq
where
$$
B_{i} \,=\,
\frac{l_{i1} + l_{i2}\,n + l_{i3}\,n^2 + l_{i4}\, n^3 + l_{i5}\,n^4}
{2\,(4\,x + y) \,( 4\,x - 4\,z + n\,y + 4\,n\,z)}\,,
\qquad\qquad i=1,2
$$
and
\beq
l_{11} &=& -16\left( 4x - y - 2z \right) \left( 2x + y + 2z
\right),
\nonumber
\\
l_{12} &=& -4\left( 16x^2 + 20xy + 5y^2 - 4xz + 6yz +
12z^2 \right),
\nonumber
\\
l_{13} &=& -16x^2 - 8xy - 10y^2 - 72xz - 56yz - 8z^2 ,\;\;
\nonumber
\\
l_{14} &=& -4xy + 2y^2 - 8xz + 6yz - 16z^2,
\qquad\qquad\quad
l_{15} = 2z\left( y + 4z \right) \,\,;
\nonumber
\\
l_{21} &=& 0
\,,\qquad\qquad
l_{22} = - 4 (2x + 2z + y)
\,,\qquad\qquad
l_{23} = 4   (3x + z + y)\,,
\nonumber
\\
l_{24} &=& y - 4x + 8z\,,
\qquad
l_{25} = - y - 4z \,\,;
\nonumber
\eeq

The computation of \ $\tr(\hat{V}^{\rh\la}R_{\rh\la})$ \ yields
\beq
\tr(\hat{V}^{\rh\la}R_{\rh\la}) = C_6\,R_{\mu\nu}^2 +
C_7\,R^2 + C_3\,\frac{R}{\ka^2}\, ,
\eeq
where
\beq C_i & = &
\frac{p_{i1} + p_{i2}\, n + p_{i3}\,n^2 + p_{i4}\, n^3} {\left(
4\,x + y \right) \,\left( 4\,x - 4\,z +
n\,\left( y + 4\,z \right)  \right)}\, , \;\;\; i = 6,7\;\; ,\nonumber \\
C_3 & = & \frac{- \left( n^2 - 3\,n + 2 \right) \,
      \left( 4\,x + n\,y + 2\,y +
        4\,z + 4\,n\,z \right) }{2\,
    \left( 4\,x + y \right) \,\left( 4\,x - 4\,z +
      n\,\left( y + 4\,z \right)  \right) }\, , \nonumber
\eeq
with
\beq
p_{61} & = & -8\,\left( 4\,x - y - 2\,z \right) \,\left( 2\,x + y + 2\,z
\right)\, , \nonumber \\
p_{62} & = & -2\,\left( 4\,x^2 + 14\,x\,y + 3\,y^2 + 20\,x\,z + 8\,y\,z +
16\,z^2 \right)\, , \nonumber \\
p_{63} & = & 8\,x^2 + 2\,x\,y - 3\,y^2 - 16\,x\,z - 16\,y\,z\, ,\;\;\;\;
p_{64} = \left( 2\,x + y \right) \,\left( y + 4\,z \right)\, , \nonumber \\
p_{71} & = & -4\,\left( 6\,x^2 + 3\,x\,y + y^2 - 12\,x\,z - y\,z - 2\,z^2
\right), \nonumber \\
p_{72} & = & -2\,\left( 8\,x^2 + 3\,x\,y + y^2 + 10\,x\,z + 6\,y\,z +
2\,z^2 \right)\, , \nonumber \\
p_{73} & = & -4\,x\,y - 12\,x\,z - y\,z - 8\,z^2\, \;\;\;\;\;\;
p_{74} = z\,\left( y + 4\,z \right)\, . \nonumber
\eeq

The results for the traces
$\frac{1}{48}\tr (\hat{V}^\la_\la \hat{V}^\rh_\rh) +
\frac{1}{24}\tr(\hat{V}_{\la\rh}\hat{V}^{\la\rh})$
can be written as
\beq
\frac{1}{48}\tr (\hat{V}^\la_\la \hat{V}^\rh_\rh) +
\frac{1}{24}\tr(\hat{V}_{\la\rh}\hat{V}^{\la\rh}) =
D_1\,R_{\mu\nu\al\be}^2 + D_2\,R_{\mu\nu}^2 + D_3\,R^2 +
D_4\,\frac{R}{\ka^2} + D_5\,\frac{1}{\ka^4}\, ,
\eeq
where
\beq
D_1 & = & \frac{q_{11} + q_{12}\, n + q_{13}\,n^2 + q_{14}\, n^3}
{96{\left( 4\,x + y \right)}^2\,\left( 4\,x - 4\,z +
n\,\left( y + 4\,z \right)  \right)}\, ,
\nonumber \\
D_i & = & \frac{q_{i1} + q_{i2}\, n + q_{i3}\,n^2 + q_{i4}\, n^3 +
q_{i5}\, n^4 + q_{i6}\, n^5 + q_{i7}\, n^6}
{96{\left( 4\,x + y \right)}^2\,\left( 4\,x - 4\,z +
n\,\left( y + 4\,z \right)  \right)^2}\, ,\;\;\; i=2,3,4,5\;\; ,
\eeq
with all non-zero $q_{ij}$ given by
\beq
q_{11} & = & 1536\,x\,\left( 2\,x^2 - 6\,x\,z - y\,z \right)\, ,
\nonumber \\
q_{12} & = & 192\,\left( 26\,x^3 + 18\,x^2\,y + 3\,x\,y^2 + 6\,x^2\,z
- 2\,x\,y\,z - y^2\,z \right)\, ,
\nonumber \\
q_{13} & = & 48\,\left( 18\,x^3 + 36\,x^2\,y + 12\,x\,y^2
+ y^3 + 78\,x^2\,z + 28\,x\,y\,z + 2\,y^2\,z \right)\, ,
\nonumber \\
q_{14} & = & 24\,{\left( 3\,x + y \right) }^2\,\left( y + 4\,z \right)
\, ,
\nonumber \\
q_{21} & = & -256\,\left( 8\,x^4 + 2\,x^2\,y^2 - 4\,x\,y^3
- y^4 + 64\,x^3\,z +
    112\,x^2\,y\,z + 20\,x\,y^2\,z
\nonumber \right.
\\
    & - & \left. 160\,x^2\,z^2 - 80\,x\,y\,z^2 -
    10\,y^2\,z^2 + 32\,x\,z^3 - 8\,z^4 \right)\, ,
\nonumber \\
q_{22} & = & 128
\left( 120 x^4 + 84x^3y + 27x^2y^2 + 6xy^3 + y^4 -
    224x^3z - 148x^2yz - 54xy^2z
\right.
\nonumber \\
& - & \left.  6y^3z - 136x^2z^2 -
    148xyz^2 - 33y^2z^2 + 80xz^3 - 12yz^3 - 32z^4 \right)\, ,
\nonumber \\
q_{23} & = & 16\left( 448x^4 + 704x^3y + 360x^2y^2 + 64xy^3
+ 15y^4 + 1024x^3z + 576x^2yz
\right.
\nonumber
\\
    & + & \left. 8xy^2z + 84y^3z -
    1344x^2z^2 - 816xyz^2 + 116y^2z^2 + 64xz^3 +
    144yz^3 + 128z^4 \right)\, ,
\nonumber \\
q_{24} & = & 8\left(
64x^4 + 448x^3y + 344x^2y^2 + 104xy^3 + 5y^4 +
    1664x^3z + 1408x^2yz
    \right. \nonumber \\
    & + & \left.
560xy^2z + 28y^3z + 320x^2z^2 + 512xyz^2
+ 8y^2z^2 - 512xz^3 - 128yz^3
    \right)\, ,
\nonumber \\
q_{25} & = & 8\left( y + 4z \right) \left( 32x^3 + 56x^2y + 28xy^2
+ 3y^3 + 192x^2z + 112xyz
\right. \nonumber \\
   & + & \left. 32xz^2 + 12y^2z
+ 8yz^2 \right)\, , \qquad\qquad
q_{26}  =  32x^2{\left( y + 4z \right) }^2\, ,
\nonumber \\
q_{31} & = &
- 128\left( 56x^4 + 32x^3y + 18x^2y^2 + 6xy^3 + y^4 -
    144x^3z - 48x^2yz
\right. \nonumber \\
    & - & \left. 2y^3z - 24xy^2z + 240x^2z^2 -
    6y^2z^2 - 80xz^3 - 16yz^3 - 8z^4 \right)\, ,
\nonumber \\
q_{32} & = & - 64\left( 68x^4 + 96x^3y + 36x^2y^2 + 13xy^3 + 3y^4
+    72x^3z - 48x^2yz + 10xy^2z
 \right. \nonumber \\
    & + & \left. 13y^3z  - 216x^2z^2 -
    16xyz^2 + 26y^2z^2 + 168xz^3 + 64yz^3
+ 36z^4 \right)\, ,
\nonumber \\
q_{33} & = & -8\left( 32x^4 + 304x^3y + 204x^2y^2 + 44xy^3 + y^4
+    1088x^3z + 784x^2yz + 280xy^2z
    \right. \nonumber \\
    & + & \left. 4y^3z - 1088x^2z^2 - 80xyz^2 - 144y^2z^2 + 192xz^3
-    336yz^3 - 160z^4 \right)\, ,
\nonumber \\
q_{34} & = & -4\left( 32x^3y + 100x^2y^2 + 32xy^3 + 7y^4 + 192x^3z
+    640x^2yz + 208xy^2z
\right. \nonumber \\
    & + & \left.  1392x^2z^2 + 68y^3z
+ 448xyz^2 + 216y^2z^2 - 544xz^3 + 64yz^3
    - 80z^4 \right)\, ,
\nonumber \\
q_{35} & = & 4\left( -4x^2y^2 - 4xy^3 + y^4 - 64x^2yz - 24xy^2z
+ 6y^3z - 144x^2z^2
\right. \nonumber \\
    & - & \left. 18y^2z^2  - 8xyz^2
- 136yz^3 - 80z^4 \right)\, ,
\nonumber \\
q_{36} & = & -4z\left( y + 4z \right)
  \left( 4xy - 2y^2 + 8xz - 9yz + 4z^2 \right)\, ,\;\;\;
q_{37} = 4z^2{\left( y + 4z \right) }^2\, ,
\nonumber \\
q_{41} & = & -128\left( 8x^3 + 4x^2y + 6xy^2 + y^3
- 80x^2z - 8xyz +    40xz^2 + 4yz^2 \right)\, ,
\nonumber \\
q_{42} & = & -32\left( 16x^3
- 8xy^2 - y^3 + 120x^2z - 2y^2z - 160xz^2
-32yz^2 - 8z^3 \right)\, ,
\nonumber \\
q_{43} & = &
16\left( 16x^3 - 8x^2y + 16xy^2 + y^3 - 176x^2z
+ 24xyz - 14y^2z  \right. \nonumber \\
& + & \left. 80xz^2 - 48yz^2 - 16z^3 \right)\, ,
\nonumber \\
q_{44} & = & 8\left( 16x^3 - 12xy^2 + y^3 + 120x^2z
- 16xyz + 16y^2z - 160xz^2 - 40z^3 \right)\, ,
\nonumber \\
q_{45} & = & 4\left( 16x^2y - 4xy^2 + y^3 + 16x^2z
- 40xyz + 18y^2z + 88yz^2 + 80z^3 \right)\, ,
\nonumber \\
q_{46} & = & -4\left( y + 4z \right)
\left( -2xy + y^2 + 5yz - 4z^2 \right)\,, \qquad\qquad
q_{47} = -4z{\left( y + 4z \right) }^2\, ,
\nonumber \\
q_{51} & = & 0 \, ,\qquad\qquad
q_{52} = 16\left( 6x + y - 2z \right) \,
\left( 2x + y + 2z \right)\, ,
\nonumber \\
q_{53} & = & -8\left( 24x^2 + 12xy + y^2
- 4yz - 8z^2 \right)\,,
\nonumber \\
q_{54} & = & -8\left( 6x^2 + 8xy + y^2 + 20xz
- 10z^2 \right)\, , \;\;\;
\nonumber \\
q_{55}  & = &
2\left( 24x^2 + 12xy - y^2 - 20yz - 40z^2 \right)\, ,
\nonumber \\
q_{56} & = & \left( 8x + y - 4z \right) \left( y + 4z \right)\,,
\qquad\qquad
q_{57}  =  {\left( y + 4z \right) }^2\, .
\nonumber
\eeq

\vskip 2mm

Finally, let us present details of the contributions from
Faddeev-Popov ghosts and weight operator (sometimes called third
ghost). The action of the ghosts has the form \beq S_{gh} \,=\,
\int d^4 x\sqrt{g}\,\,{\bar C}^\mu \,\left({\cal
H}_{gh}\right)_\mu^\nu\,C_\nu\, , \label{ghost action 1} \eeq
where the operator \ $\hat{\cal H}_{gh}$ \ is given by the
expression
$$
\hat{\cal H}_{gh} \,=\, ({\cal H}_{gh})^\al_\be
\,=\, \frac{\de\ch^\al}{\de h^{\mu\nu}}
\,{\cal D}^{\mu\nu\,,}\mbox{}_\be\,.
$$
Here $\,{\cal D}^{\mu\nu\, ,}\mbox{}_\be\,$ is generator of
the diffeomorphism transformations for the metric \
$\de g^{\mu\nu} = {\cal D}^{\mu\nu\,,}\mbox{}_\be\,\de x^\be$
$$
{\cal D}_{\mu\nu\, ,}\mbox{}_\be
\,=\, - g_{\mu\be}\,\na_\nu - g_{\nu\be}\,\na_\mu\,.
$$
By direct computation one derives, assuming $p_1 = p_2 = 0$ in
(\ref{fixing}) and keeping other gauge-fixing parameters arbitrary
(\ref{ghost operator})
\beq
\hat{\cal H}_{gh}=\left({\cal
H}_{gh}\right)_\mu^\nu = - \de_\mu^\nu\,\Box - \na^\nu \na_\mu -
2\be \na_\mu \na^\nu\,.
\label{ghost operator 0}
\eeq
In order to
evaluate the expression \ $- i\,\ln\,\Det\hat{\cal{H}}_{gh}$, let
us rewrite (\ref{ghost operator 0}) in the form suitable for the
generalized Schwinger-DeWitt method \cite{bavi85}
\beq
{\hat{\cal
H}}_{gh}= - \left[\de_\mu^\nu\,\Box - \si\, \na_\mu\na^\nu  +
P_{\mu}^{\nu}\right]\,,
\label{ghost operator2}
\eeq
where \ $\si = -(1+2\be)$ \ and \ $P_{\mu\nu} = R_{\mu\nu}$.
The \ $n$ \
dimensional analog of the known algorithm \cite{frts82,bavi85} for
the non-minimal Abelian vector operator (\ref{ghost operator2})
has the form
\beq
- \frac{i}{2}\,\ln\,\Det\hat{\cal{H}}_{gh}|_{{\rm div}}
& = &
\frac{\mu^{n-4}}{(4\pi)^2(n-4)} \int d^n
x\sqrt{g}\,\left\{\frac{n-15}{180} R_{\mu\nu\al\be}^2
+ \Big( \frac{1}{24}\psi^2 + \frac{1}{12}\psi -\frac{n}{180}
\Big)R_{\mu\nu}^2 \right. \nonumber
\\
& + & \left. \Big(\frac{1}{48}\psi^2 + \frac{1}{12}\psi
+ \frac{n}{72} \Big)R^2 + \Big( \frac{1}{12}\psi^2
+ \frac{1}{3}\psi \Big) R_{\mu\nu}P^{\mu\nu}
\right.
\nonumber
\\
& + & \left.
\Big( \frac{1}{24}\psi^2 + \frac{1}{4}\psi + \frac{1}{2}
\Big)P_{\mu\nu}^2
+ \Big( \frac{1}{24}\psi^2 + \frac{1}{12}\psi
+ \frac{1}{6} \Big) R P + \frac{1}{48} \psi^2 P^2
\right\}\,,
\label{nonminimaldet}
\eeq
where \ $P = P_\al^\al$ \ and
$$
\psi = \frac{\si}{1 - \si} = -\frac{1+2\be}{2+2\be}\,.
$$
Taking into account $P_{\mu\nu} = R_{\mu\nu}$, we find
\beq
- i\,\ln\,\Det\hat{\cal{H}}_{gh}|_{{\rm div}} & = &
\frac{\mu^{n-4}}{(4\pi)^2(n-4)}\,\int d^n x\sqrt{g}\,
\left\{\frac{n-15}{90} R_{\mu\nu\al\be}^2 \right.
\nonumber
\\
& + & \left. \Big(\frac{1}{3}\psi^2 + \frac43\psi
+ \frac{90-n}{90}\Big) R_{\al\be}^2 + \Big( \frac{1}{6}\psi^2
+ \frac13\psi + \frac{n + 12}{36}\Big) R^2\right\}\,.
\eeq
The
contribution of the weight operator (\ref{fixing}) is calculated
with the choice (\ref{minimal}). The divergent part of $\Tr\ln
\hat{Y}$ can be computed directly from (\ref{nonminimaldet}). As
usual \cite{frts82,bavi85}, in the case \ $P_{\mu\nu} = -
R_{\mu\nu}$, \ the gauge dependence is canceled (this
cancellation is related to the gauge invariance of the
electromagnetic field, whose contribution is given by exactly
the same operator):
\beq
- \frac{i}{2}\,\ln\,\Det\hat{Y}|_{{\rm div}}
= \frac{\mu^{n-4}}{(4\pi)^2(n-4)}\,\int d^n x\sqrt{g}\,
\left\{\frac{n-15}{180} R_{\mu\nu\al\be}^2 + \frac{90-n}{180}
R_{\al\be}^2 + \frac{n - 12}{72} R^2\right\}.
\label{weight
contribution}
\eeq

\section*{\large\bf
Appendix C. Coefficients of $\be$-functions in $4-\ep$
dimension}

This Appendix contains the coefficients $\de$ of the
$\be$-functions (\ref{beta's}), most of them were not
included into the main text. The total expressions
for the $\be$-functions is as follows (\ref{beta's})
\beq
(4\pi)^2\,\be_1(\rh\,,\la\,,\xi\,,\ep)
= \de_{1A}^{(0)} + \de_{1B}^{(0)}\,\frac{\la}{\xi}
- \de_{1A}^{(1)}\,\frac{\xi}{\rho}
- \de_{1B}^{(1)}\,\frac{\la}{\rho}
+ \de_{1A}^{(2)}\,\frac{\xi^2}{\rho^2} +
\de_{1B}^{(2)}\,\frac{\la\,\xi}{\rho^2} +
\de_{1C}^{(2)}\,\frac{\la^2}{\rho^2}
 \, ,\nonumber
\eeq
\beq
(4\pi)^2\,\be_2(\rh\,,\la\,,\xi\,,\ep) =
\de_{2A}^{(0)}\,\frac{\xi}{\la} + \de_{2B}^{(0)} +
\de_{2C}^{(0)}\,\frac{\la}{\xi} - \de_{2A}^{(1)}\,\frac{\xi}{\rho}
- \de_{2B}^{(1)}\,\frac{\la}{\rho} +
\de_{2A}^{(2)}\,\frac{\xi^2}{\rho^2} +
\de_{2B}^{(2)}\,\frac{\la\,\xi}{\rho^2} + \de_{2C}^{(2)}\,
\frac{\la^2}{\rho^2}\, , \nonumber
\eeq
\beq
(4\pi)^2\,\be_3(\rh\,,\la\,,\xi\,,\ep) & = & \de_{3A}^{(0)}+
\de_{3B}^{(0)}\,\frac{\la}{\xi}+ \de_{3C}^{(0)}\,
\frac{\la^2}{\xi^2}-\de_{3A}^{(1)}\,\frac{\xi}{\rho} -
\de_{3B}^{(1)}\,\frac{\la}{\rho}-
\de_{3C}^{(1)}\,\frac{\la^2}{\xi\,\rho}
 \nonumber \\
& + & \de_{3A}^{(2)}\,\frac{\xi^2}{\rho^2}
+ \de_{3B}^{(2)}\,\frac{\la\,\xi}{\rho^2} +
\de_{3C}^{(2)}\,\frac{\la^2}{\rho^2}\, .
\label{betas}
\eeq
The coefficients \ $\de_{1\,A,B,C}^{(i)}$ \ have the form
\beq
\de_{1A}^{(0)} & = & \frac{37632 - 92704\,\epsilon  + 81860\,{\epsilon }^2 -
    31350\,{\epsilon }^3 + 5533\,{\epsilon }^4 -
    446\,{\epsilon }^5 + 15\,{\epsilon }^6}{2880\,
    \left( \ep -3  \right) \,
    {\left( \ep -1  \right) }^2} \, ,
\nonumber \\
\de_{1B}^{(0)} & = &
-\frac{\ep\,\left( \ep -6  \right) \,
      {\left( \ep -2  \right) }^2 }{48\,
    {\left( \ep -1  \right) }^2}\, ,
\nonumber \\
\de_{1A}^{(1)} & = &
-\frac{\epsilon \,\left( \ep + 4  \right) }
  {12\,{\left( \ep - 3  \right) }^2}\, ,
\nonumber \\
\de_{1B}^{(1)} & = &
\frac{\ep\,\left( \ep - 2  \right) \,
    \left( 92 - 176\,\epsilon  + 57\,{\epsilon }^2 -
      10\,{\epsilon }^3 + {\epsilon }^4 \right) }{48\,
    \left( \ep - 3  \right) \,
    {\left( \ep - 1  \right) }^2}\, ,
\nonumber \\
\de_{1A}^{(2)} & = &
-\frac{\ep^2\,\left( \ep - 2  \right) \,
      \left( \ep - 1  \right) }{96\,
    {\left( \ep - 3  \right) }^4}\, ,
\nonumber \\
\de_{1B}^{(2)} & = &
-\frac{\ep^2\,\left( \epsilon - 5  \right) \,
      \left( \ep - 4  \right) \,
      \left( \ep - 2  \right) }{48\,
    {\left( \ep - 3  \right) }^3}\, ,
\nonumber \\
\de_{1C}^{(2)} & = &
\frac{\ep^2\,\left( \ep - 5  \right) \,
    {\left( \ep - 2  \right) }^2\,
    \left( {\epsilon }^2 - 12\,\ep + 29 \right) }{48\,
    {\left( \ep - 3  \right) }^2\,
    {\left( \ep - 1  \right) }^2}\, .
\eeq

For the expressions \ $\de_{2\,A,B,C}^{(i)}$\ , we have
\beq
\de_{2A}^{(0)} & = &
-\frac{ \epsilon \,\left( \ep^2 +
    4\,\ep - 4 \right)}{32\,
    {\left( \ep - 3  \right) }^2\,
    \left( \ep - 1  \right) }\, ,
\nonumber \\
\de_{2B}^{(0)} & = &
-\frac{ \left( \ep - 2 \right) \,
      \left( -9576 + 21180\,\epsilon  - 15410\,{\epsilon }^2 +
        4410\,{\epsilon }^3 - 399\,{\epsilon }^4
      + 5\,{\epsilon }^5 \right)}{960\,\left( \ep - 3\right)\,
    {\left( \ep - 1 \right) }^2}\, ,
\nonumber \\
\de_{2C}^{(0)} & = &
\frac{\ep\,\left( \ep - 6  \right) \,
    {\left( \ep - 2  \right) }^2}{48\,
    {\left( \ep - 1  \right) }^2}\, ,
\nonumber \\
\de_{2A}^{(1)} & = &
\frac{\ep\,\left( \ep - 2  \right) \,
    \left( 2\,\epsilon + 1  \right) }{24\,
    {\left( \ep - 3  \right) }^2\,
    \left( \ep - 1 \right) }\, ,
\nonumber \\
\de_{2B}^{(1)} & = &
-\frac{\ep\,\left( \epsilon - 7  \right) \,
      {\left( \ep - 2  \right) }^2\,
      \left({\epsilon }^2 - 4\,\epsilon + 10\right) }
    {48\,\left( \ep - 3  \right) \,
    {\left( \ep - 1  \right) }^2}\, ,
\nonumber \\
\de_{2A}^{(2)} & = &
\frac{\ep^2\,\left( \epsilon - 2  \right)
\,\left( \epsilon - 1  \right)}
{96\,{\left( \epsilon - 3 \right) }^4}\, ,
\nonumber \\
\de_{2B}^{(2)} & = &
\frac{\ep^2\,{\left( \ep - 2  \right) }^2\,
    \left( {\epsilon }^2 - 8\,\epsilon + 19 \right) }{48\,
    {\left( \ep - 3  \right) }^3\,
    \left( \ep - 1 \right) }\, ,
\nonumber \\
\de_{2C}^{(2)} & = &
-\frac{ \ep\,{\left( \epsilon - 2 \right) }^3\,
      \left( -162 + 95\,\epsilon  - 18\,{\epsilon }^2 +
        {\epsilon }^3 \right) }{48\,
    {\left( \ep - 3  \right) }^2\,
    {\left( \ep - 1  \right) }^2}\, .
\eeq

\vskip 1mm

The coefficients $\de_{3\,A,B,C}^{(i)}$ can be written as
\beq
\de_{3A}^{(0)} & = &
\frac{-1200 + 8808\,\epsilon  - 12480\,{\epsilon }^2 +
    5980\,{\epsilon }^3 - 1290\,{\epsilon }^4 + 127\,{\epsilon }^5 -
    5\,{\epsilon }^6}{960\,{\left( \epsilon - 3  \right) }^2\,
    \left( \epsilon - 1 \right) }\, ,
\nonumber \\
\de_{3B}^{(0)} & = &
-\frac{{\left( \epsilon - 2  \right) }^2\,
      \left( 180 + 28\,\epsilon  - 16\,{\epsilon }^2 +
        {\epsilon }^3 \right)}{48\,
    \left( \epsilon - 3  \right) \,\left( \epsilon - 1  \right) }\, ,
\nonumber \\
\de_{3C}^{(0)} & = &
\frac{\left( \ep - 6  \right) \,
    \left( \ep - 5  \right) \,\left( \ep - 4  \right) \,
    {\left( \epsilon - 2  \right) }^3}{96\,
    {\left( \epsilon - 1  \right) }^2}\, ,
\nonumber \\
\de_{3A}^{(1)} & = &
\frac{\ep\,\left( \epsilon - 2 \right) \,
    \left( \epsilon - 1  \right)\,
    \left( {\epsilon }^2 - 20 \right) }{96\,
    {\left(\epsilon - 3  \right) }^3}\, ,
\nonumber \\
\de_{3B}^{(1)} & = &
-\frac{ {\ep\,\left( \epsilon - 2  \right) }^2\,
      \left( -10 + 24\,\epsilon  - 9\,{\epsilon }^2 +
        {\epsilon }^3 \right) }{24\,
    {\left( \epsilon - 3  \right) }^2\,
    \left( \epsilon - 1  \right) }\, ,
\nonumber \\
\de_{3C}^{(1)} & = &
\frac{\ep\,\left( \epsilon - 6  \right) \,
    \left( \epsilon - 5  \right) \,
    {\left( \epsilon - 2  \right) }^3}{48\,
    \left( \epsilon - 3  \right) \,\left( \epsilon - 1  \right) }\, ,
\nonumber \\
\de_{3A}^{(2)} & = &
\frac{\ep^2\,\left(\epsilon - 5  \right) \,
    {\left( \epsilon - 2 \right) }^2\,
    {\left( \epsilon - 1  \right) }^2} {192\,
    {\left( \epsilon - 3 \right) }^5}\, ,
\nonumber \\
\de_{3B}^{(2)} & = &
\frac{\ep^2\,\left( \epsilon - 5 \right) \,
    {\left( \epsilon - 2  \right) }^2\,
    \left( \epsilon - 1  \right)}{48\,
    {\left( \epsilon - 3  \right) }^4}\, ,
\nonumber \\
\de_{3C}^{(2)} & = &
\frac{\ep^2\,\left( \epsilon - 5 \right) \,
    {\left( \epsilon - 2 \right) }^3\,
    \left( {\epsilon }^2 - 8\,\epsilon + 19 \right) }{96\,
    {\left( \epsilon - 3  \right) }^3\,
    \left( \epsilon - 1 \right) }\, .
\eeq

The coefficients $\de_{4\,A,B,C}^{(i)}$ are the following:
\beq
\de_{4A}^{(0)} & = &
-\frac{\left( \epsilon - 2  \right) \,
      \left( 72 - 40\,\epsilon  - 4\,{\epsilon }^2 +
        {\epsilon }^3 \right)}{192\,
    {\left( \epsilon - 3  \right) }^2}\, ,
\nonumber \\
\de_{4B}^{(0)} & = &
\frac{{\left( \epsilon - 2  \right) }^2\,
    \left( -156 + 144\,\epsilon  - 27\,{\epsilon }^2 +
      {\epsilon }^3 \right) }{96\,\left( \epsilon - 3 \right) \,
    \left( \epsilon - 1  \right) }\, ,
\nonumber \\
\de_{4C}^{(0)} & = &
-\frac{\left( \epsilon - 6  \right) \,
      \left( \epsilon - 5 \right) \,
      \left( \epsilon - 4  \right) \,
      {\left( \epsilon - 2  \right) }^3 }{96\,
    {\left( \epsilon - 1  \right) }^2}\, ,
\nonumber \\
\de_{4A}^{(1)} & = &
-\frac{\ep\,\left( \epsilon - 6  \right) \,
      {\left( \epsilon - 2 \right) }^2\,
      \left( \epsilon - 1 \right) }{192\,
    {\left( \epsilon - 3 \right) }^3}\, ,
\nonumber \\
\de_{4B}^{(1)} & = &
-\frac{\ep\,\left( \epsilon - 6  \right) \,
      \left( \epsilon - 5  \right) \,
      {\left( \epsilon - 2 \right) }^3}{96\,
    \left( \epsilon - 3 \right) \,\left( \epsilon - 1 \right) }\, ,
\eeq

The remaining coefficients corresponding
to $\de_{5\,A,B,C}^{(i)}$ and $\de_{6\,A,B,C}^{(i)}$ have the form
\beq
\de_{5A}^{(0)} & = &
-\frac{\left( \epsilon - 4  \right) \,
      \left( \epsilon - 2 \right) }{4\,
    {\left( \epsilon - 3 \right) }^2}\, ,
\qquad
\qquad\quad
\de_{5B}^{(0)}  =
\frac{\left( \epsilon - 4  \right) \,
    \left( \epsilon - 2  \right) \,
    \left(  {\epsilon }^2 - 8\,\epsilon + 14 \right) }{2\,
    \left( \epsilon - 3 \right) \,\left( \epsilon - 1  \right) }\, ,
\nonumber
\\
\de_{6A}^{(0)} & = &
\frac{\left( \epsilon - 6 \right) \,
    \left( \epsilon - 4  \right) \,
    {\left( \epsilon - 2 \right) }^2}{768\,
    {\left( \epsilon - 3  \right) }^2}\, ,
\qquad
\de_{6B}^{(0)} =
\frac{\left( \epsilon - 6  \right) \,
    \left( \epsilon - 5 \right) \,\left( \epsilon - 4  \right) \,
    {\left( \epsilon - 2 \right) }^3}{384\,
    {\left( \epsilon - 1 \right) }^2}\, .
\eeq

\section*{\large\bf  Appendix D. Renormalization group
equations at \ $n=4$}

For convenience we present the solution
\cite{frts82,avrbar}
of the $n=4$  renormalization group equations (\ref{betas 4}).
The first two equations may be solved immediately.
\beq
\rho(t)=\frac{\rho_0}{1+b^2\rho_0\,t}\,,
\qquad \rho_0 = \rho(0)\,,
\qquad b^2 = \frac{196}{45\,(4\pi)^2}\,;
\nonumber
\\
\nonumber
\\
\la(t)=\frac{\la_0}{1+a^2\la_0\,t}\,,
\qquad
\la_0 = \la(0)\,,
\qquad
a^2 = \frac{133}{10\,(4\pi)^2}\,.
\label{a f 1}
\eeq
In the UV limit $t\to \infty$ the solutions (\ref{a f 1})
manifest the asymptotic freedom for the two parameters.
Indeed, the term ``asymptotic freedom'' is proper only for
$\,\la$, which is a parameter of the loop expansion of
theory (\ref{action}). The positivity in the theory
is consistent with \ $\la > 0$ \ and therefore, according
to (\ref{a f 1}), we need $\,\la_0>0$  \ too. In
contrast, $\,\rho^{-1}$ \ is the coefficient of the topological
Gauss-Bonnet term, which does not correspond to any
gravitons interaction in the $n=4$ case.
The third equation (\ref{betas 4}) can be rewritten in
terms of a new  variable \ $w(t)=\xi(t)/\la(t)$
\beq
\Big(t+\frac{1}{a^2\la_0}\Big)
\,\frac{d w}{dt}
= \frac{C}{a^2}\,\left(w^2 - kw + l\right)
= \frac{C}{a^2}\,\left(w-w_1\right)\,\left(w-w_2\right)\,,
\label{w}
\eeq
where
\beq
C=\frac{10}{(4\pi)^2}\,,
\qquad
k = \frac{183}{133}\,,
\qquad
l = \frac{25}{133 \cdot 18}\,.
\qquad
w_{1/2}=\frac{k}{2}\pm\sqrt{\frac{k^2}{4}-l}\,.
\label{w1}
\eeq
Let us notice that $w_1\approx k \gg l/k \approx w_2 >0$.
The solution of Eq. (\ref{w}) can be easily found and
written in terms of the original variables, notations
(\ref{w1}) and \ $w_0=w(0)$, \ as
\beq
w\,=\,\frac{w_1-X\,w_2}{1-X}
\,,\qquad \mbox{where}\qquad
X\,=\,
\left(\,\frac{w_0-w_1}{w_0-w_2}\,\right)\,\cdot\,
\left( 1 + \la_0 a^2t\right)^m
\label{w2}
\eeq
and
\beq
m\,=\,\frac{C\,(w_1-w_2)}{a^2}\,\approx\,0.517\,.
\nonumber
\eeq
The UV behavior of \ $w(t)$ \ depends on \ $w_0$.

i) For \ $w_0 < w_2$ \ the UV limit is \ $w \to w_2$;

ii) $w_0 = w_{1,2}$ \ corresponds to the fixed points
 \ $w \equiv w_{1,2}$;

iii) For \ $w_2 < w_0 < w_1$ \ the UV limit is \ $w \to w_2$,
\ hence $w_2$ is a stable fixed point of the theory;

iv) For \ $w_0 > w_1$ \ the UV limit is singular.
The singularity is achieved at \ $X=1$, that
corresponds to the value
\beq
t_s = \frac{1}{\la_0\,a^2}\,
\left[ \Big(\frac{w_0-w_1}{w_0-w_2}\Big)^{1/m} \,-\, 1\right]\,.
\label{w3}
\eeq
For $\,w_0$ \ comparable to \ $w_1$ \ $t_s \gg 10 \la_0^{-1}$.
We remark that the applicability of the perturbative
approach requires small $\,\la_0$ \ and therefore the
singularity occurs at very high energies. For $\,w_0 \gg w_1\,$
the position of the singularity point \ $t_s$ \ may be closer to

zero. Since the value \ $t_s$ \ is finite, this point is singular
also for \ $\xi$.

\vskip 20mm



\begin{thebibliography}{99}

\bibitem{hove}
G. t'Hooft and M. Veltman, Ann. Inst. H. Poincare. {\bf A20},
69 (1974).

\bibitem{dene} S.Deser and P. van Nieuwenhuisen,  Phys. Rev.
      {\bf D10}, 401 (1974); {\bf D10}, 411 (1974).

\bibitem{stelle} K.S. Stelle, Phys.Rev. {\bf D16}, 953 (1977).

\bibitem{frts82} E.S. Fradkin and  A.A. Tseytlin,
Nucl. Phys. {\bf 201B} (1982) 469.

\bibitem{Weinberg79} S. Weinberg,
in: {\it General Relativity.}
ed: S.W. Hawking and W. Israel (Cambridge. Univ. Press. 1979).

\bibitem{ambjorn}
J. Ambjorn, J. Jurkiewicz and R. Loll,
Nucl. Phys. {\bf 610B} (2001) 347, see also further references
therein;

Herbert W. Hamber, Phys. Rev. {\bf D61} (2000) 124008.

\bibitem{smolin} L. Smolin,
{\sl An invitation to loop quantum gravity,} [hep-th/0408048],
see further references therein.

\bibitem{reuter}
O. Lauscher and M. Reuter, Phys. Rev. {\bf D66} (2002) 025026;
{\bf D65} (2002) 025013.

\bibitem{GorSag}
M.H. Goroff and A. Sagnotti, Nucl. Phys. {\bf 266B}, 709 (1986).

\bibitem{west} P. West, {\it Introduction to Supersymmetry and
Supergravity} (World Scientific, 1986).

\bibitem{votyu84} B.L. Voronov and I.V. Tyutin,
{\sl Sov. J. Nucl. Phys.} {\bf 39} (1984) 998.

\bibitem{julton}
J. Julve and M. Tonin, Nuovo Cim. {\bf 46B} (1978) 137.

\bibitem{salstr}  A. Salam and J. Strathdee,
{\sl Phys. Rev.} {\bf D18} (1978) 4480.

\bibitem{avrbar}
I.G. Avramidi, A.O. Barvinsky, Phys. Lett. {\bf 159B} (1985) 269;

I.G. Avramidi, Sov. J. Nucl. Phys. {\bf 44} (1986) 255;

I.G. Avramidi,
{\sl Covariant methods for the calculation of the effective action
in quantum field theory and investigation of higher-derivative
quantum gravity} (Ph.D. thesis, Moscow University, 1986);
 [hep-th/9510140].

\bibitem{bush}
I.L. Buchbinder and I.L. Shapiro, Sov. J. Nucl. Phys. {\bf 44},
(1986) 1033;

I.L. Buchbinder, O.K. Kalashnikov, I.L. Shapiro,
V.B. Vologodsky and Yu.Yu. Wolfengaut, Phys.  Lett.
{\bf 216B}, (1989) 127;

I.L. Shapiro, Class. Quant. Grav. {\bf 6}, (1989) 1197.

\bibitem{highderi}
M. Asorey, J.L. L\'opez and I.L. Shapiro,
Int. Journ. Mod. Phys. {\bf 12A} (1997) 5711. 

\bibitem{book} I.L. Buchbinder, S.D. Odintsov and I.L. Shapiro,
 {\sl Effective Action in Quantum Gravity} (IOP, Bristol, 1992).

\bibitem{accioly}
A. Bartoli, J. Julve and E.J. Sanchez,
Class. Quant. Grav. {\bf 16} (1999) 2283. 

A. Accioly, A. Azeredo and H. Mukai, Journ. Math. Phys. {\bf 43}
(2002) 473. 

\bibitem{nevill}
D.E. Nevill,
Phys.Rev. {\bf D18} (1978) 3535;  {\bf D21} (1980) 867.

\bibitem{sezgin} E. Sezgin and P. van Nieuwenhuizen,
Phys. Rev. {\bf D21} (1981) 3269;

E. Sezgin, Phys. Rev. {\bf D24} (1981) 1677.

\bibitem{tyutin} I.V. Tyutin, private conversation.

\bibitem{torsi} I.L. Shapiro,
Phys. Repts. {\bf 357} (2002) 113; [hep-th/0103093].

\bibitem{tom}
Tomboulis E., {\sl Phys. Lett.} {\bf 70B} (1977) 361.

\bibitem{anto}
E.T. Tomboulis,
Phys. Rev. Lett. {\bf 52} (1984) 1173;
Phys. Lett. {\bf 97B} (1980) 77;

I. Antoniadis and E.T. Tomboulis,
Phys. Rev. {\bf D33} (1986) 2756.

\bibitem{dgo}
D.A. Johnston, Nucl. Phys. {\bf 297B} (1988) 721.

\bibitem{Wilson71}
K. G. Wilson and M. E. Fisher,
Phys. Rev. Lett. {\bf 28} (1972) 240;

see also K.G. Wilson and J.B. Kogut,
Phys. Rept. {\bf 12} (1974) 75.

\bibitem{Wilson73}
K. G. Wilson, Phys. Rev. {\bf D7} (1973) 2911.

\bibitem{GKT} R. Gastmans, R. Kallosh and C. Truffin,
Nucl. Phys. {\bf 133B} (1978) 417.

\bibitem{ChrDuff}
S.M. Christensen and M.J. Duff, Phys. Lett. {\bf 79B} (1978) 213.

\bibitem{kawai} H. Kawai and M. Ninomiya, Nucl. Phys.
{\bf 336B} (1990) 115;


H. Kawai, Y. Kitazawa and M. Ninomiya, Nucl. Phys. {\bf 404B}
(1993) 684; {\bf 393B} (1993) 280.

\bibitem{sakai} S. Kojima, N. Sakai and Y. Tanii,
{\sl Nucl. Phys.} {\bf 426B} 223 (1994).

\bibitem{capkim} D.M. Capper and D. Kimber,
Journ. Phys. {\bf A13} (1980) 3671.

\bibitem{Weyl} G. de Berredo-Peixoto and I.L. Shapiro,
Phys. Rev. {\bf D70} (2004) 044024; hep-th/0307030.

\bibitem{antmot} I. Antoniadis, P.O. Mazur and E. Mottola,
Nucl. Phys. {\bf 388B} (1992) 627.

\bibitem{GSW}
M.B. Green, J.H. Schwarz and E. Witten,
{\it Superstring Theory}
(Cambridge University Press, Cambridge, 1987).

\bibitem{don}
J. Donoghue, Phys. Rev. Lett. {\bf 72} (1994) 2996;
Phys. Rev. {\bf D50} (1994) 3874;

N.E.J Bjerrum-Bohr, J.F. Donoghue and B.R. Holstein,
Phys. Rev. {\bf D67} (2003) 084033.

\bibitem{tsamis}
N.C. Tsamis and  R.P. Woodard, Ann. Phys. {\bf 238} (1995) 1,
see further references therein.

\bibitem{bavi85} A.O. Barvinsky and G.A. Vilkovisky,
Phys. Repts. {\bf 119} (1985) 1.

\bibitem{Christen} N.H. Barth and S.M. Christensen,
Phys. Rev. {\bf D28} (1983) 1876.

\bibitem{shja} I.L. Shapiro and A.G. Jacksenaev,
Phys. Lett. {\bf 324B} (1994) 286.

\bibitem{VLT} B.L. Voronov,  P.M. Lavrov and I.V. Tyutin,
 {\sl Sov. J. Nucl. Phys.} {\bf 36} (1982) 498.

\bibitem{applications}
O. Bertolami, J.M. Mourao and J. Perez-Mercader,
Phys. Lett. {\bf 311B} (1993) 27;

T. Goldman, J. Perez-Mercader, F. Cooper and M.M. Nieto,
Phys. Lett. {\bf 281B} (1992) 219;

O. Bertolami and  F.M. Nunes, Phys. Lett. {\bf 452B} (1999) 108;

A. Bonanno and M. Reuter, Phys. Rev. {\bf D65} (2002) 043508;

O. Bertolami and J. Garcia-Bellido
Int. J. Mod. Phys. {\bf D5} (1996) 363.

\bibitem{MAPLE} A. Heck,
{\it Introduction to Maple}, 
(Springer-Verlag, 2003).

\bibitem{M4} S. Wolfram,
{\it The Mathematica Book, Fifth Edition}
(Cambridge University Press, 1996).

\bibitem{MathTensor}
L. Parker and S.M. Christensen,
{\it MathTensor: A System for Doing Tensor Analysis
by Computer} (Addison-Wesley P.C., 1994).

\bibitem{Polyakov}  A.M. Polyakov,
{\sl Int. J. Mod. Phys.} {\bf 16} (2001) 4511, and references
therein.

\bibitem{cosm} I. Shapiro and J. Sol\`{a}, Nucl. Phys.
(Proceedings Supplement) {127B} (2004) 71;

I.L. Shapiro, J. Sol\`a,
C. Espa\~{n}a-Bonet and P. Ruiz-Lapuente,
Phys. Lett. {\bf B574} (2003) 149.

\bibitem{apco}
E.V. Gorbar and I.L. Shapiro,   JHEP {\bf 02} (2003) 021;
JHEP {\bf 06} (2003) 004.

\end{thebibliography}
\end{document}